\documentclass{aa}
\usepackage{soul}
\usepackage{graphicx} 
\usepackage{txfonts}
\usepackage{natbib}
\usepackage{hyperref}
\usepackage{siunitx}
\usepackage{xcolor}
\def\modif#1{{#1}}

\graphicspath{
	{/home/petri/publications/preparations/matteo/}
}

\begin{document}
\title{The Galactic population of canonical pulsars II}
\author{Mattéo Sautron\inst{1} \and Jérôme Pétri\inst{1} \and Dipanjan Mitra\inst{2,}\inst{3} \and Ludmilla Dirson\inst{1}}
\institute{Université de Strasbourg, CNRS, Observatoire astronomique de Strasbourg, UMR 7550, 67000 Strasbourg, France \and National Centre for Radio Astrophysics, Tata Institute for Fundamental Research, Post Bag 3, Ganeshkhind, Pune 411007, India \and Janusz Gil Institute of Astronomy, University of Zielona Góra, ul. Szafrana 2, 65-516 Zielona Góra, Poland}
\date{Received / Accepted}

\titlerunning{The Galactic population of canonical pulsars - II}
\authorrunning{Sautron et al.}

\abstract{
Pulsars are highly magnetized rotating neutron stars, emitting in a broad electromagnetic energy range. These objects have been discovered more than 55~years ago and are astrophysical laboratories for studying physics at extreme conditions. Reproducing the observed pulsars population refines our understanding of their formation and evolution scenarios as well as their radiation processes and geometry.}
{In this paper, we improve our previous population synthesis by focusing on both the radio and $\gamma$-ray pulsar populations, investigating the impact of the Galactic gravitational potential and of the radio emission death line. 
In order to  elucidate the necessity of a death line, refined initial distributions of spin period and spacial position at birth were implemented, elevating the sophistication of our simulations to the most recent state-of-the-art.}
{The motion of each individual pulsar is tracked in the Galactic potential by a fourth order symplectic integration scheme. Our pulsar population synthesis takes into account the secular evolution of the force-free magnetosphere and magnetic field decay simultaneously and self-consistently. Each pulsar is evolved from its birth up to the present time. The radio and $\gamma$-ray emission locations are modelled respectively by the polar cap geometry and the striped wind model.}{
By simulating ten million pulsars we found that including a death line better reproduces the observational trend. However, when simulating one million pulsars, we obtain an even more realistic $P-\dot{P}$ diagram, whether or not a death line is included. This suggests that the ages of the detected pulsars might be overestimated, therefore questioning the real need for a death line in pulsar population studies. Kolmogorov-Smirnov tests confirm the statistical similarity between the observed and simulated $P-\dot{P}$ diagram. Additionally, simulations with increased $\gamma$-ray telescope sensitivities hint to a significant contribution of $\gamma$-ray pulsars to the GeV excess in the Galactic centre.}
{}


\maketitle
\nolinenumbers
\section{Introduction}
Pulsars, discovered by Jocelyn Bell Burnell in 1967 \citep{BellHewish} are rotating neutron stars born in a core-collapse supernova. They are highly magnetized and surrounded by a plasma-filled magnetosphere emitting regular pulses of radiation at their spin frequency. Due to the magneto-dipole losses, they lose rotational kinetic energy and their spin period increases. Extensive radio surveys, performed for instance by the Parkes and Arecibo radio-telescopes,  so far 3700 pulsars have been discovered (\citealt{Manchester}, see Table~\ref{tabl_nbpulsar_known} listing the canonical pulsar population \modif{defined by isolated pulsars with periods longer than 20 msec and that are not magnetars}). Although pulsars were first observed in radio, they were later found to be also bright in X-rays, optical and $\gamma$-rays. The Large Area Telescope (LAT) on board of the Fermi satellite, has discovered dozens of radio-quiet $\gamma$-ray pulsars as well as millisecond pulsars (MSPs). Since its launch, the LAT has detected about 300~pulsars \citep{Abdo2013,Smith2019,Smith2023} and unveiled a new perspective, expanding our understanding about pulsar emission mechanisms by increasing the sample of neutron stars detectable through their high energy emission. Moreover, the distinct radiation process involved in high-energy photon production and its unique beaming properties offer an alternative viewpoint on the global pulsar population.
\begin{table}[h]
\caption{Number of known canonical pulsars, with spindown luminosity~$\dot{E}$ above and below $10^{28}$~W, and above $10^{31}$~W. The quantities $N_{\rm r}$, $N_{\rm g}$, and $N_{\rm rg}$ are the number of radio-only, gamma only, and radio-loud $\gamma$-ray pulsars, respectively. It should be noted that we excluded the binary pulsars, the pulsars in the Magellanic clouds and in globular clusters, \modif{then we selected the pulsars with a magnetic field between $6\times10^6$~T and $4.4\times10^9$~T.} The data have been taken from the ATNF catalogue from \href{https://www.atnf.csiro.au/research/pulsar/psrcat/}{https://www.atnf.csiro.au/research/pulsar/psrcat/}}
\label{tabl_nbpulsar_known} 
\centering 
\begin{tabular}{c c c c c} 
\hline\hline 
log($\dot{E}$) (W) & $N_{\rm tot}$ & $N_{\rm r}$ & $N_{\rm g}$ & $N_{\rm rg}$\\
\hline 
>31 & 4 & 0 & 2 & 2\\ 
>28 & 150 & 40 & 44 & 66\\
Total & 2193 & 2020 & 80 & 93\\
\hline 
\end{tabular}
\end{table}

With the advent of future surveys such as the Square Kilometre Array (SKA \footnote{\href{https://www.skao.int/en/science-users/118/ska-telescope-specifications}{https://www.skao.int/en/science-users/118/ska-telescope-specifications}}) or the Cherenkov Telescope Array (CTA \footnote{\href{https://www.cta-observatory.org/about/how-ctao-works/}{https://www.cta-observatory.org/about/how-ctao-works/}}), much more pulsars will be detected in both these wavelengths. Those instruments are scheduled to start collecting data by the end of the decade. SKA is going to be 50~times more sensitive than current telescopes and will survey the sky 10,000~times faster than any existing imaging radio telescope \citep{SKApaper}. CTA is aimed at detecting $\gamma$-rays in the energy range from a few tens of GeV up to hundreds of TeV \citep{CTApaper} while Fermi/LAT operates in the energy range from 20~MeV to several hundred GeV \footnote{\href{https://fermi.gsfc.nasa.gov/science/instruments/table1-1.html}{https://fermi.gsfc.nasa.gov/science/instruments/table1-1.html}} and H.E.S.S operates in the energy range from 0.03 to 100 TeV \footnote{\href{https://www.mpi-hd.mpg.de/HESS/pages/about/}{https://www.mpi-hd.mpg.de/HESS/pages/about/}}.

Pulsar population synthesis (PPS) is a powerful tool to predict the discovery rate of new pulsars, to better understand their emission processes and to constrain their overall properties. In a PPS, pulsars are generated from their birth and evolved up to the present time. Once the sample of pulsars is generated, detection criteria are applied to check whether they can be detected by current and future telescopes. The detectability of an individual pulsar in a given energy band, radio or $\gamma$-ray, depends on whether the associated emission beam intersects our line of sight and on the sensitivity of the corresponding instrument. 

Most PPS studies assume that neutron stars are rotating in vacuum \citep{FaucherG,Popov,Johnston20} or take only radio or $\gamma$-ray emission into account separately \citep{Watters2011,Gullon14}, or assume a constant magnetic field during the evolution process \citep{Gonthier,FaucherG,Johnston2017}. 
\citet{Dirson22} have modelled for the first time the population of both the $\gamma$-ray and radio pulsars, taking into account the state-of-the-art force-free magnetosphere model in conjunction with a prescription for the magnetic field decay. The neutron star radiation is produced by relativistic charged particles flowing within their magnetosphere. Therefore the plasma back reaction must be taken into account in the evolution of the pulsar period and magnetic inclination angle. 
Whereas the location of the $\gamma$-ray emission site is still under debate, different regions have been suggested. These are for instance the polar cap region \citep{Sturrock,Ruderman,DaughertyHarding82,DaughertyHarding96} at low altitudes, the slot gap along the last open magnetic field lines \citep{Arons,Muslimov,Harding,HardingMuslimov}, or the outer gaps at high altitudes within the light-cylinder, in the outer magnetosphere \citep{Cheng1,Hirotani,Takata}. Emission is also possible outside the light cylinder, in the so-called striped wind \citep{Petri2009,Petri2011} and was used for the first time by \citet{Dirson22} in a PPS study. The radio emission is described as usual by the polar cap model. In this work as in \citet{Dirson22} only the canonical pulsar population in our Galaxy is reproduced. 
 
While the study by \citet{Dirson22} yielded satisfactory results when simulating a population closely resembling the observed pulsar population in the $P-\dot{P}$ diagram, there was room for improvement regarding the pulsar trajectories within the Milky Way. Specifically, \citet{Dirson22} study was limited by the assumption that pulsars move in a ballistic motion depicted by straight lines at a constant speed from their birth to the present time. A noticeable discrepancy arises when comparing the observed position of pulsars to the predictions, revealing a significant disparity in both spatial distributions.

\citet{Dirson22} did not require a death line to reproduce the $P-\dot{P}$diagram and was therefore not implemented in their PPS. However the death line segregates neutron stars generating particles through pair cascading from those not radiating any more because of the quenching of this pair avalanche, and hence they are called dead pulsars. The effect of the death line assumption on the PPS differs from the PPS performed by \citet{Graber} who did not consider a death line but instead relied only on as special radio luminosity law to eliminate some pulsars in the $P-\dot{P}$ diagram. 

In this paper we further improve the PPS of \citet{Dirson22} by examining the necessity of a death line to explain the observed canonical pulsars population and discuss in depth the $\gamma$-ray pulsar population. Moreover, our study helps to better understand the impact of changing the parameters of our model as well as their physically meaningful ranges. The paper is organised as follows. In Sect.~\ref{S1} we detail our PPS model, recalling the processes for generation and evolution of the pulsar sample, the description of the Galactic potential and of the death line, discussing our adopted multi-wavelength detection criteria. The results are shown in Sect.~\ref{S4}. The results are discussed in Sect.~\ref{S5} and Sect.~\ref{S6} covers a summary of the results.

\section{Description of the PPS model} \label{S1}

Our PPS study is based on a self-consistent state-of-the-art evolution of the neutron star geometry of radio and $\gamma$-ray emission, magnetic field and proper motion within the Milky Way. In this section we detail our model by exposing, first the generation of individual pulsars, their initial position, magnetic field and spin period, second how these quantities evolve in time. We also give explicit expressions for the Galactic potential used in our simulations and introduce the death line before discussing the radio and $\gamma$-ray detection criteria. We stress that the model parameters used in this section have been carefully chosen by comparing simulated data with actual observations, varying the unknown input parameters in sensible physical ranges (see appendix~\ref{AppB} for more details). 

\subsection{Generation of pulsars} \label{S11}
We generate a population of pulsars whose real ages are selected in ascending order, for example every X~years if the birth rate is 1/X~yr$^{-1}$, with their age being between 0~yr and the age of the Milky Way at maximum. The age of the Milky Way does not significantly differ from the age of the Universe estimated about $t_H \approx 13.8 \times 10^9$~yr, therefore, a total number of $t_H / X$ pulsars should be simulated. The best value for this birth rate was found to be 1/41~yr$^{-1}$, see appendix~\ref{AppB}. Our value is consistent with earlier investigations by \cite{FaucherG,Gullon14,Johnston2017} where they evaluated the birth rate of pulsars to be in the range [1/150,1/33]~yr$^{-1}$. With $X=41$~yr we should generate more than $10^7$ pulsars, however we chose to emulate only $10^7$ pulsars in a first run and $10^6$ in a second run for the following two reasons. Firstly, the computation time is significantly decreased, secondly pulsars older than $10^7$~yr are mostly not detected. This approach was also chosen because it allows us to verify whether our birth rate is realistic, by comparing our results from the simulation to the observations. Our approach is different from previous works like \citet{Gullon14} and \citet{Johnston2017}, where they stop the simulation whenever their detected number of pulsars equals the number of observed pulsars. 

To describe the position of the pulsars in the Galaxy, we use the right-handed Galacto-centric coordinate system $(x,y,z)$ with the Galactic centre at its origin, $y$ increasing in the disc plane towards the location of the Sun, and $z$ increasing towards the direction of the north Galactic pole. The initial spatial distribution of the pulsars is given by \citet{Pacz}. He found two distributions, one describing the radial spread and one the spread in altitude, however only his distribution in altitude is used in our work because the radial distribution in the galactic plane is better depicted in  \citet{Yusifov} by the Milky Way's pulsar surface density defined as
\begin{subequations}
\begin{align}
\label{Yusu_eq}
\rho(R) &=A\left(\frac{R+R_1}{R_{\odot}+R_1}\right)^a \exp\left(-b\left(\frac{R-R_{\odot}}{R_{\odot}+R_1}\right)\right)  \\
\label{Pacz_eq_2}
\rho_z(z) & = \frac{e^{-|z|/h_c}}{h_c}       
\end{align}
\end{subequations}
where $R$ is the axial distance from the $z$-axis, and $z$ is the distance
from the Galactic disc. The numerical values for the constants are, $A = 37.6$~kpc$^{-2}$, $a = 1.64$, $b=4.0$, $R_1=0.55$~kpc, $h_c=180$~pc and $R_{\odot}=8.5$~kpc and in agreement with the distribution of young massive stars in our galaxy \citep{Lietal}.
The purpose of choosing equation \eqref{Yusu_eq} to describe the radial spread is to put neutron stars birth positions within the spiral arms of the Galaxy. More precisely the Galactic spiral structure contains four arms with a logarithmic shape function allowing to obtain the azimuthal coordinate $\phi$ as a function of the distance from the Galactic center
\begin{equation}
\label{spiral_arms_eq}
\phi(R) = k \ln\left(\frac{R}{r_0}\right) +\phi_0 . 
\end{equation}
The values of the models describing each arm are given in Table~\ref{tabl_param_MW_spiral_arms}. These values were used by \citet{Ronchi} and taken from \citet{Yaoetal} to match the shape of the arms of the Galaxy. The local arm is not modeled, because of its very low density, much smaller than the four other arms. Each star has an equal probability to be in one of the four arms, its angular coordinate~$\phi$ for a given $R$ being deduced from \eqref{Yusu_eq}. Furthermore, the Galaxy is not static, and its arms are moving with an approximated period $T=250$~Myr \citep{Skowron}. Since we also know that the Galaxy is rotating in the clockwise direction, by knowing the age of a pulsar we can know its angular position at birth. Following the same procedures as \citet{Ronchi}, in order to avoid artificial features near the Galactic center, noise is added to both coordinates R and $\phi$. For instance $\phi_{corr}=\phi_{rand} \, \exp(-0.35\,R)$ where $\phi_{rand}$ is randomly drawn from a uniform distribution between $0$ and $2\,\pi$. $r_{corr}$ will be taken from a normal distribution with a mean of $0$ and a standard deviation $\sigma_{corr}=0.07\,R$ and we add this two corrections to $\phi$ and $R$ respectively. Therefore $R_{birth}=R+r_{corr}$ and $\phi_{birth}=\phi+\phi_{corr}+ \frac{2\,\pi\, t_{age}}{T}$, where $t_{age}$ is the age of the pulsar. As soon as $R_{birth}$ and $\phi_{birth}$ are known we can convert these coordinates in $x$ and $y$ Galacto-centric coordinates for each pulsar. 
\begin{table}[h]
\caption{Parameters of the Milky Way Spiral Arm structure, adapted from \citet{Yaoetal}.} 
\label{tabl_param_MW_spiral_arms} 
\centering 
\begin{tabular}{c c c c c} 
\hline\hline
Arm Number & Name & \(k \) & \( r_0 \) & \( \phi_0 \) \\
& & (rad) & (kpc) & (rad) \\
\hline 
1 & Norma & 4.95 & 3.35 & 0.77  \\ 
2 & Carina-Sagittarius & 5.46 & 3.56 & 3.82 \\ 
3 & Perseus & 5.77 & 3.71 & 2.09  \\ 
4 & Crux-Scutum & 5.37 & 3.67 & 5.76  \\ 
\hline 
\end{tabular}
\end{table} 

Next, we need to describe the individual properties of each pulsar, namely, its inclination angle $\alpha$, which is the angle between its rotation axis and its magnetic axis, its initial spin period~$P_0$ and initial magnetic field~$B_0$ at birth. The inclination angle $\alpha$ is assumed to follow an isotropic distribution generated from a uniform distribution $U\in[0,1]$ and given by $\alpha = \arccos (2\,U-1)$.
In our work, the initial spin period and magnetic field follow both a log-normal distribution, as suggested from the results of a study of 56~young neutron stars by \citet{Igo} while other works simply use a Gaussian distribution for the spin period at birth \citep{FaucherG,Gullon14,Johnston20} while for the magnetic field the prescription is the same as ours.
Explicitly, the probability distributions for an initial magnetic field $B_0$ and an initial period $P_0$ are given by
\begin{gather}
p(\log(B_0)) = \frac{1}{\sigma_b \sqrt{2\pi}} e^{-(\log B_0 - \log \bar{B})^2/(2\sigma_b^2)} \\
p(\log(P_0)) = \frac{1}{\sigma_p \sqrt{2\pi}} e^{-(\log P_0 - \log \bar{P})^2/(2\sigma_p^2)} \ 
\end{gather}
The distribution for the magnetic field is identical to the one used in other pulsar population synthesis like for instance \citet{FaucherG}, \citet{Gullon14}, \citet{Johnston20}, \citet{Yadi} and \citet{Watters2011}. 
 
According to \citet{Hobbs} a Maxwellian distribution for the kick velocity at birth best replicates the observations and is given by
\begin{equation} \label{eq:maxwellian_speed}
p(v) = \sqrt{\frac{2}{\pi}} \frac{v^2}{\sigma_v^3} \exp{\left(-\frac{v^2}{2\sigma_v^2}\right)}
\end{equation}
The mean velocity of the distribution is related to the standard deviation by $ \bar{v} = \sigma_{\rm v} \sqrt{{8}/{\pi}}$ with a standard deviation $\sigma_{\rm v} = 265$~km/s for pulsars. The velocity value drawn from the Maxwellian distribution is then distributed along the unit vector of the rotation axis which is generated, as detailed in subsection \ref{S13}. Moreover compared to the study of \citet{Dirson22} we introduce a novelty, namely the alignment between the kick velocities at birth and the rotation axis of the pulsar, as suggested by \citet{Rankin}.

\subsection{Pulsar evolution} \label{S12}

The pulsar initial period~$P_0$, inclination angle~$\alpha_0$ and magnetic field~$B_0$ are evolved in time in a fully self-consistent way taking into account the spin-down losses and the internal magnetic field dissipation within the neutron star. The neutron star period is evolved according to the force-free magnetosphere model which takes into account the electric current and charge flowing inside the magnetosphere. The spin down luminosity~$\dot{E}$ is 
\begin{gather}
\dot{E} = \frac{dE_{\rm rot}}{dt} = - I \, \Omega \, \dot{\Omega} = L_{\rm ffe} \label{eq:spin_down_eq} \\ 
L_{\rm ffe} = \frac{4\pi R^6 B^2 \Omega^4(1+\sin^2\alpha)}{\mu_0 c^3} \label{eq:value_of_spin_down_eq}
\end{gather} 
where $I$ is the neutron star moment of inertia approximately equal to $I \approx 10^{38}$~kg.m$^2$, $\Omega = 2\pi / P$ the rotation frequency of the pulsar ($P$ is the spin period), $\dot{\Omega}$ its time derivative, $R = 12$~km the typical radius of a neutron star as found by recent NICER observations \citep{Riley,Bogdanov}, $B$ the magnetic field at the magnetic equator, $\alpha$ the inclination angle, $\mu_0$ the vacuum permeability constant, $\mu_0=4 \pi \times 10^{-7}$ H/m and $c$ the speed of light. The term $L_{\rm ffe}$ for the spin down luminosity was given by \citet{Spitkovsky2006} and \citet{Petri2012}. Equation~\eqref{eq:spin_down_eq} clearly shows the correlation between the obliquity, the magnetic field and the rotation frequency. Combining equation \eqref{eq:spin_down_eq} and \eqref{eq:value_of_spin_down_eq} leads to 
\begin{gather}
\dot{\Omega} = -K_{\rm ffe} \ \Omega^n \label{eq:general_omegadot} \\ 
K_{\rm ffe} = \frac{4\pi R^6 B^2(1+\sin^2\alpha)}{\mu_0 c^3 I} \label{eq:kffe}
\end{gather}
which is written in a more concise form where $n$ is the braking index, its value being $n=3$ for magnetic dipole radiation.

The integral of motion between $\Omega$ and $\alpha$ is also very useful and reads
\begin{equation} \label{eq:integralofmotion}
\Omega \frac{\cos^2\alpha}{\sin\alpha} = \Omega_0 \frac{\cos^2\alpha_0}{\sin\alpha_0}
\end{equation}
where quantities with subscript~0 indicate their initial value and those without a subscript their current value at present time. 

As shown by \citet{Gullon14} a decaying magnetic field better accounts for comparison between pulsar population synthesis and observations. Therefore, we prescribe a magnetic field decay according to a power law 
\begin{equation} \label{eq:Bfield}
B(t) = B_0(1+t/\tau_d)^{-1/\alpha_d}
\end{equation}
where $\alpha_d$ is a constant parameter controlling the rate of the magnetic field decay and $\tau_d$ the typical decay timescale, which depends on the initial magnetic field as demonstrated by magneto-thermal evolution models \citep{Viganoetal}. We took these results into account by defining the decay timescale $\tau_1$ for a magnetic field strength $B_1$ such that $\tau_1 \, B_1^{\alpha_d} = \tau_d \, B_0^{\alpha_d}$. Thus, $\tau_1$ is another magnetic field decay time scale, corresponding to a certain magnetic field value $B_1$. Three different $\tau_1$ are considered in this work and are randomly chosen for each pulsar generated : $1.5\times10^5$~yr or $3.5\times10^5$~yr or $2.5\times10^6$~yr with each of these timescales \modif{that} corresponds to a magnetic field value $B_1$ of $1\times10^8$~T, $3\times10^8$~T and $2\times10^9$~T. Then the probabilities to get these values of $\tau_1$ and $B_1$ to be adopted for one pulsar are 0.23, 0.46 and 0.31 respectively. \modif{These three probabilities actually approximate the probability density function for $\tau_1$ and $B_1$, and were already sufficient to reproduce the canonical population. Finding the complete distribution for $\tau_1$ and $B_1$ would require further investigations not needed in our study. Moreover, the distributions of $\tau_1$ and $B_1$ would physically represent the whole ensemble of trajectories possible in the $P-\dot{P}$ diagram for canonical pulsars as in Fig. 10 of \citet{Viganoetal}. Therefore, each pulsar has three possible evolutionary paths in the simulation, which is an empirical way to reproduce the $P-\dot{P}$ diagram in a better way than having only one evolutionary path. These particular choices for the discrete values were only guided by trial and errors.} The value of $\alpha_d$ can be found in Table \ref{tabl_sim_para} of Sect. \ref{S4}. 

In line with the spin down luminosity, the inclination angle satisfies another evolution equation that after integration was found by \citet{Phillipov14} for a spherically symmetric neutron star with a constant magnetic field. 
With our decaying prescription the inclination angle $\alpha$ is found by solving
\begin{align} 
\label{eq:incl_angle}
\begin{split}
\ln(\sin \alpha_0) + \frac{1}{2\sin^2\alpha_0} + K \Omega_0^2 \frac{\cos^4\alpha_0}{\sin^2\alpha_0} \frac{\alpha_d \tau_d B_0^2}{\alpha_d - 2} \left[\left(1+\frac{t}{\tau_d}\right)^{1-2/\alpha_d}-1\right] \\ 
\phantom{{}={}}= \ln(\sin \alpha) + \frac{1}{2\sin^2\alpha}
\end{split}
\end{align}
where $t$ is the time representing the age of the pulsar and $K={R^6}/{I\,c^3}$. The typical decay timescale for a mean magnetic field of $2.5 \times 10^8$~T is $4.6 \times 10^5$~yr in \citet{Vigano}, however in this study the typical decay timescale associated to this magnetic field is $3.5\times 10^5$~yr which is very close to the estimate of \citet{Vigano} for the same field strength.

\subsection{Description of the Galactic potential} \label{S2}

Finally, we need to follow the particle motion within the Galactic potential. Each pulsar evolves in the gravitational potential~$\Phi$ subject to an acceleration $\vec{\ddot{x}}$ according to 
\begin{equation} \label{move_equation}
\vec{\ddot{x}} = -\vec{\nabla} \Phi \ 
\end{equation}
Equation~\eqref{move_equation} is integrated numerically thanks to a Position Extended Forest Ruth-Like (PEFRL) algorithm \citep{PEFRL}, a fourth order integration scheme, presented in Appendix \ref{AppC} with a convergence and accuracy study. We now discuss the galactic potential model used.

The Galaxy is divided in four distinct regions with different mass distributions and associated gravitational potentials. The four potentials are: the bulge~$\Phi_{b}$, the disk~$\Phi_{d}$, the dark matter halo~$\Phi_{h}$ and the nucleus~$\Phi_{n}$. The total potential of the Milky Way~$\Phi_{tot}$ is the sum of these potentials
\begin{equation} \label{tot_pot}
\Phi_{tot} = \Phi_{b} + \Phi_{d} + \Phi_{h} + \Phi_{n} .
\end{equation} 
The expressions for these potentials are taken from \citet{BB21} with parameters given in Table~\ref{tabl_const_pot}. The nucleus mass was found in \citet{Bovy}. 
\begin{table}[h]
\caption{Constants values used for the different potentials, with the solar mass $M_{\odot}=1.99 \times 10^{30}$ kg.} 
\label{tabl_const_pot} 
\centering 
\begin{tabular}{c c} 
\hline\hline 
Parameters & Values \\
\hline 
$M_h$ & $2.9 \times 10^{11} \pm 7.6 \times 10^{10} M_{\odot}$  \\ 
$a_h$ & 7.7 $\pm$ 2.1 kpc \\
$a_d$ & 4.4 $\pm$ 0.73 kpc \\
$b_d$ & 0.308 $\pm$ 0.005 kpc  \\
$M_d$ & $6.50 \times 10^{10} \pm 1.9 \times 10^9  M_{\odot}$ \\
$a_b$ & 0.0 kpc \\
$b_b$ & 0.267 $\pm$ 0.009 kpc \\
$M_b$ & $1.02 \times 10^{10} \pm 6.3 \times 10^8 M_{\odot}$ \\
$M_{n}$ & $4\times10^6 \pm 0.42 \times 10^6 M_{\odot}$ \\
\hline 
\end{tabular}
\end{table} 

The potential for the bulge $\Phi_b$ and for the disk $\Phi_d$ were both chosen to have the form proposed by \citet{MNpot} which are typically used for models of the gravitational potential of the Milky Way. They read
\begin{equation} \label{potMN}
\Phi_i(R,z) = -\frac{GM_i}{\left[R^2+\left(a_i+\sqrt{z^2+b_i^2}\right)^2\right]^{1/2}}
\end{equation}
where $R^2=x^2+y^2$, and $i=b$ is for the bulge and $i=d$ is for the disk. In those formula $R$ depends on the coordinates $x$ and $y$, $a_i$ and $b_i$ are the scale parameters of the components in kpc, $M_i$ is the mass (for the disk or the bulge) and $G$ is the gravitational constant. The constant values can be found in Table \ref{tabl_const_pot}.

Solving the pulsar equation of motion requires to compute the gradient of the potentials, the derivatives of the disk and the bulge potentials are
\begin{equation} \label{deriv_pot_MN1}
\frac{\partial \Phi_i}{\partial x_j} = \frac{G M_i x_j}{\left[R^2+\left(a_i+\sqrt{z^2+b_i^2}\right)^2\right]^{3/2}}
\end{equation}
\begin{equation} \label{deriv_pot_MN2}
\frac{\partial \Phi_i}{\partial z} = \frac{G M_i z\left(a_i+\left(z^2+b_i^2\right)^{\frac{1}{2}}\right)}{\left[R^2+\left(a_i+\sqrt{z^2+b_i^2}\right)^2\right]^{3/2} \ \left(z^2+b_i^2\right)^{1/2}}
\end{equation}
$x_j$ being either $x$ or $y$.
Equation~\eqref{deriv_pot_MN1} is the derivative for the potential of Miyamoto Nagai, where $i=d$ or $i=b$ depending on disk or bulge derivative. Equation~\eqref{deriv_pot_MN2} is the derivative of the Miyamoto Nagai potential with respect to the $z$ coordinate. 

The potential for the dark matter halo $\Phi_h$, according to the frequently used potential of \citet{NFW}, is expressed as
\begin{equation} \label{NFW_eq}
\Phi_h(r)=-\frac{GM_h}{r} \ \ln\left(1+\frac{r}{a_h}\right)
\end{equation}
where $M_h$ is the mass of the halo, $r=x^2+y^2+z^2$, and $a_h$ is the length scale whose values are found in Table~\ref{tabl_const_pot}. The derivative of equation~\eqref{NFW_eq} with respect to any coordinate $x_j$ $(x,y,z)$ is
\begin{equation} \label{deriv_NFW}
\frac{\partial \Phi_h}{\partial x_j} = \frac{G M_h x_j}{r^2} \ \left[\frac{1}{r} \ \ln\left(1+\frac{r}{a_h}\right) - \frac{1}{r+a_h}\right] .
\end{equation}

Finally for the last part of the Galaxy, the nucleus of the Milky Way is simply represented by a Keplerian potential $\Phi_{n}$ such as
\begin{equation} \label{kep_pot}
\Phi_{n}(r) = -\frac{GM_{n}}{r}
\end{equation} 
where $M_{n}$ is the mass of the nucleus. The derivative of equation \eqref{kep_pot} with respect to any coordinate $x_j \ (x,y,z)$ is given by
\begin{equation} \label{deriv_keppot}
\frac{\partial \Phi_{n}}{\partial x_j} = \frac{GM_{n} \, x_j}{r^3} .
\end{equation}

\subsection{The death line} \label{DL_sub}


Several equations for the death line have been proposed in the literature \citep{ChenRuderman93,Zhangetal,GilMitra,Mitraetal}. 
The most suitable one in our study is that of \citet{Mitraetal} where they 
found an expression relating $P$ and $\dot{P}$ by
\begin{equation}\label{death_line}
\dot{P}_{\rm line} = \frac{3.16 \times 10^{-19} \ T_6^4 \ P^2}{\eta^2 b \cos^2 \alpha_l} \ .
\end{equation}
Here $T_6 = T/10^6$~K, where $T$ corresponds to the surface temperature of the polar cap. The parameter $\eta = 1 - \rho_i/\rho_{GJ}$ is the electric potential screening factor due to the ion flow, where $\rho_i$ correspond to the ion charge density and $\rho_{GJ}$ to the Goldreich-Julian charge density above the polar cap.
The quantity $b$ is the ratio of the actual surface magnetic field to the dipolar surface magnetic field and $\alpha_l$ is the angle between the local magnetic field and the rotation axis. 
The parameters we used for this death line are : $\alpha_l=45\degr$, $b=40$, $\eta=0.15$, $T_6=2$. Spread in the parameter values of the model causes significant variations in the death line, and thus a death valley rather than a single death line describes the condition for pulsar extinction in the $P-\dot{P}$ diagram.

\begin{figure}[h]
\resizebox{\hsize}{!}{\includegraphics{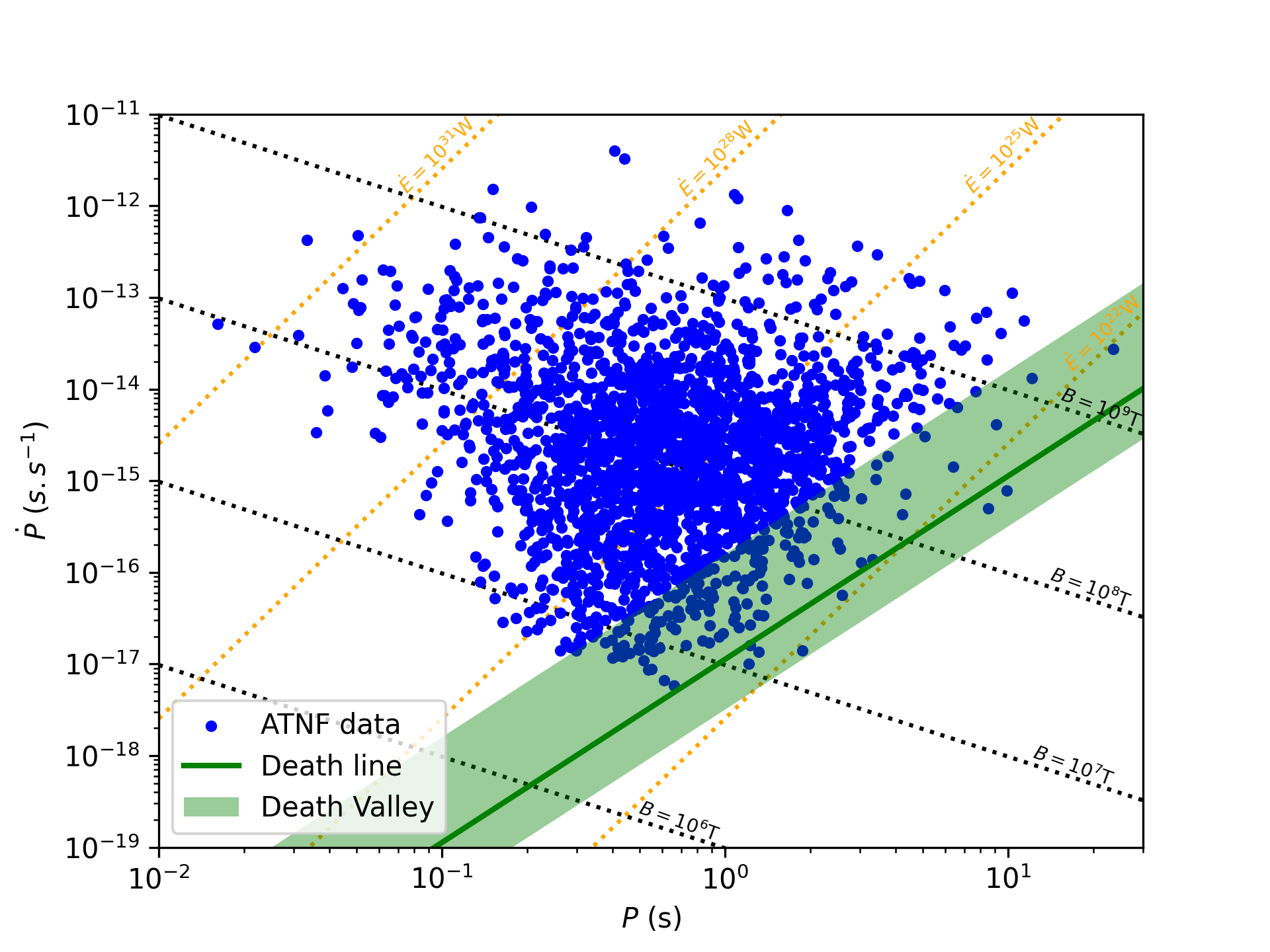}}
\caption{$P-\dot{P}$ diagram of the canonical pulsars along with the death line, green solid line, and death valley, shaded green area.}
\label{Pulsar_obs_all}
\end{figure}

We allow the different parameters for this death valley to be in the following ranges : $\alpha_l$ is drawn from a uniform distribution between $0\degr$ and $65\degr$, $T_6$ is drawn from a uniform distribution between 1.9 and 2.8 and $b$ is drawn from a uniform distribution between 30 and 60. The boundary parameter values, corresponding to the edges of the death valley, were chosen to traverse the $P-\dot{P}$ diagram from the region in the bottom right quadrant where the number of pulsars starts to decrease (because this is likely the region where the death of pulsars begins to be considered, since less pulsars are there), to the detectable pulsar with the lowest $\dot{P}$ and highest $P$ in the same quadrant, as shown in Fig. \ref{Pulsar_obs_all}. 
To decide whether a pulsar is still active in radio or not we apply the following criterion: if the pulsar lies in the death valley 
then all the parameters mentioned above are taken from uniform distributions in the ranges of the death valley in order to compute $\dot{P}^{\rm new}_{\rm line}$ with equation \eqref{death_line}, the critical spin down rate of the considered pulsar. Then we compare its actual spin down rate $\dot{P}$, computed from the evolution model, to the $\dot{P}^{\rm new }_{\rm line}$. For pulsars that are not in the death valley, their $\dot{P}$ is only compared with the spin down rate $\dot{P}_{\rm line}$ obtained with the default parameters.
If the pulsar has its $\dot{P}$, which was computed from the evolution model, below the one computed with the death line,
then the pulsar is considered as dead, otherwise it is considered as still emitting photons. These choices were made in order to have more like a death valley near the death line than a strict condition of death, because the parameters $T_6$, $\alpha_l$ and $b$ may be different from one pulsar to another. 

\subsection{Detection} \label{S13}

For each pulsar, we check whether it fulfills the detection criteria depending on three factors. First, the beaming fraction indicates the fraction of the sky covered by the radiation beam and depends on the considered wavelength, here radio or $\gamma$-ray. The beaming fraction varies also with the pulsar spin rate, geometry and location of the emission regions. Before explaining how the beaming fraction in radio and in $\gamma$-ray are computed, it is important to define several angles. 

The angle between the line of sight and the rotation axis is denoted by $\xi = (\widehat{\vec{n_{\Omega}},\vec{n}})$ where $\vec{n}_{\Omega} = {\vec{\Omega}}/{\|\vec{\Omega}\|}$ is a unit vector along the rotation axis and $\vec{n}$ the unit vector along the line of sight. The inclination angle $\alpha=(\widehat{\vec{n_{\Omega}},\vec{\mu}})$ is the angle between the rotation axis and the magnetic moment, $\vec{\mu}$ being the unit vector along the magnetic moment. Traditionally the impact angle is also introduced as the angle between the magnetic moment and the line of sight $\beta = (\widehat{\vec{\mu},\vec{n}})$. Moreover, it is related to the previous angles by $\alpha + \beta = \xi$.

We choose an isotropic distribution for the Earth viewing angle $\xi$ as well as for the orientation of the unitary rotation vector. The Cartesian coordinates of the unit rotation vector $\vec{n_{\Omega}}$ are $(\sin\theta_{n_{\Omega}} \cos\phi_{n_{\Omega}}, \sin\theta_{n_{\Omega}} \sin\phi_{n_{\Omega}}, \cos\theta_{n_{\Omega}})$. We set the Sun position at $(x_{\odot}, y_{\odot}, z_{\odot}) =$ (0~kpc, 8.5~kpc, 15~pc) \citep{Siegert}. The coordinates for $\vec{n}$ are
\begin{equation} \label{eq:coord_n_vector}
\vec{n} = \left(\frac{x-x_{\odot}}{d},\frac{y-y_{\odot}}{d},\frac{z-z_{\odot}}{d} \right) .
\end{equation}
To compute the pulsar distance from Earth, we use the formula for the distance which is
\begin{equation} \label{eq:dist}
d = \sqrt{(x-x_{\odot})^2 + (y-y_{\odot})^2 + (z-z_{\odot})^2 } .
\end{equation}

\subsubsection{Radio detection model}
The beaming fraction in radio depends on the half opening angle of the radio emission cone~$\rho$ computed according to \citet{LorimerKramer} by
\begin{equation} \label{eq:rhoangle}
\rho = 3 \sqrt{\frac{\pi \, h_{\rm em}}{2 P c}}
\end{equation}
where $h_{\rm em}$ is the emission height, $P$ is the spin period of the pulsar and $c$ is the speed of light. The emission height is taken constant with an average value of \modif{$h_{\rm em} = 3\times10^5$} m, estimated from observations of a sample of pulsars by \cite{WeltevredeJohnston,Mitra2017,Johnston2019,Johnston2023}. The cone half-opening angle $\rho$ estimated in eq.~\eqref{eq:rhoangle} holds only for the last open field lines of a magnetic dipole and can only be applied for slow pulsars where the radio emission altitude is high enough for the multipolar components to decrease significantly and become negligible.

The pulsar is detected in radio if $\beta = |\xi - \alpha| \leq \rho$ or if $\beta = |\xi - (\pi - \alpha)| \leq \rho$ corresponding to the north and south hemisphere respectively. It must also satisfy the condition $\alpha \geq \rho$ and $\alpha \leq \pi - \rho$ in order to effectively see radio pulsation, because the line of sight must cross the emission cone to observe pulsations. Another useful quantity is the observed width of the radio profile $w_r$ which is computed as (see \citet{LorimerKramer})
\begin{equation} \label{eq:observedwidthprofile}
\cos(\rho) = \cos(\alpha)\cos(\xi) + \sin(\alpha)\sin(\xi)\cos(w_r/2) .
\end{equation}

The second factor for detection is the luminosity, which is also different between radio and $\gamma$-rays. First, concerning the radio flux density, the formula used is the same as in \citet{Johnston20} for a pulsar at 1.4 GHz, in order to model the detection carried out by the Parkes radio telescope in the southern Galactic plane \citep{Kramer2003,Lorimer2006,Cameron2020} and the Arecibo telescope in the northern plane \citep{Cordes2006} (note that these two telescopes do not cover all the sky in the southern and northern plane, but we did not implemented any filter in the code)
\begin{equation} \label{eq:rad_lum}
F_r = 9 \ \text{mJy} \left(\frac{d}{1 \ \text{kpc}}\right)^{-2} \left(\frac{\dot{E}}{10^{29} \ \text{W}}\right)^{1/4} \times 10^{F_{\rm j}}
\end{equation}
where $d$ is the distance in kpc and $F_{\rm j}$ is the scatter term which is modeled as a Gaussian with a mean of 0.0 and a variance of $\sigma$ = 0.2. The detection threshold in radio is set by the signal to noise ratio defined by
\begin{equation} \label{eq:SNratio}
S/N = \frac{F_r}{S_{\rm survey}^{\rm min}} .
\end{equation} 
The pulsar is detected if the signal to noise ratio $S/N$ is greater or equal to 10. $S_{\rm survey}^{\rm min}$ is the minimum flux which is related to the period of rotation $P$, the width profile of radio emission $w_r$ and the sensitivity of the survey which is the last factor for detection, and will be detailed later. We directly compute the radio flux, without computing the luminosity in radio which overlooks the fact that the luminosity received will depend on the geometry of the beam.

The third and last factor is the sensitivity, depending on the survey, therefore on the instrument used and on the pulse profile observed. A pulsar is detected in radio if the signal to noise ratio is greater than 10, but this ratio depends on the minimum flux, another function of the instrumental sensitivity. With the aim of computing the observed width pulse profile, we use the same formula as \citet{Cordes_Mc}
\begin{subequations} 
\begin{align}
&\Tilde{w}_r = \sqrt{\left( {w_r\,P}/{2\pi}\right)^2 + \tau_{samp}^2 + \tau_{DM}^2 + \tau_{scat}^2} \label{eq:pulse_profile_ISM_scatt} \\
&S_{\text{survey}}^{\text{min}} = S_0 \sqrt{\frac{\Tilde{w}_r}{P-\Tilde{w}_r}} \label{eq:sensitivity}
\end{align}
\end{subequations}
Equation~\eqref{eq:pulse_profile_ISM_scatt} takes into account density fluctuations in the interstellar medium (ISM), the dispersion and scattering during the propagation of the radio pulse when interacting with the free electrons. Moreover, the instrumental effect is also taken into account with $\tau_{samp}$, which is the sampling time of the instrument. All this mechanisms lead to a broader pulse profile. In order to determine the influence of the ISM through $\tau_{DM}$, the formula of \citet{Batesetal} is used
\modif{ \begin{equation} \label{tau_dm_eq}
\tau_{DM}=\frac{e^2 \Delta f_{ch} \, DM}{4\,\pi^2\,\epsilon_0 \, m_e \, c \, f^3} = 8.3\times10^{15} ~s \, \frac{\Delta f_{ch}}{f^3} \, DM
\end{equation}}
$e$ is the electron charge, $m_e$ its mass, $\Delta f_{ch}$ the width of frequency of the instrument channel \modif{in Hz}, $f$ the observing frequency \modif{in Hz} and $DM$ the dispersion measure \modif{in pc/cm$^3$}. Furthermore, in order to determine the influence of scattering by an inhomogeneous and turbulent ISM through $\tau_{scat}$, the empirical fit relationship from \citet{Krishnakumar} is used
\begin{equation} \label{tau_scat_eq}
\tau_{scat} = 3.6\times 10^{-9} DM^{2.2} (1+1.94\times 10^{-3} DM^2) \ .
\end{equation}
Both $\tau_{DM}$ and $\tau_{scat}$ \modif{are given in units of second (s) and depend on the dispersion measure DM (in units of pc/cm$^3$)}, which is found for each pulsar by running the code of \citet{Yaoetal} converting the distance of a pulsar into the dispersion measure thanks to their state-of-the art model of the Galactic electron density distribution. The radio survey parameters used are the ones from the Parkes Multibeam Pulsar Survey (PMPS) taken from \citet{Manchester2001}, see Table \ref{PMPS_param}. 
\begin{table}[h]
\caption{Survey parameters of the Parks Multibeam Pulsar Survey (PMPS).} 
\label{PMPS_param} 
\centering 
\begin{tabular}{c c c} 
\hline\hline 
$\Delta f_{ch}$ & $f$ & $\tau_{samp}$ \\
(kHz) & (GHz) & (µs) \\  
\hline 
3000 & 1.374 & 250 \\
\hline 
\end{tabular}
\end{table} 

Concerning equation \eqref{eq:sensitivity}, $S_0$ represents the survey parameters. \citet{Johnston20} estimated that $S_{0}$ should approximately be equal to 0.05~mJy to have a signal to noise ratio greater than 10 for normal pulsars in this survey. We chose equation \eqref{eq:sensitivity} to compute the minimum flux, to decide whether a pulsar will be detected in radio or not, however it does not reproduce the whole complexity of this quantity which could be computed more precisely, (see equation (24) of \citet{FaucherG} for instance, where $S_0$ is computed more precisely) even though it remains a good approximation. In addition we only use the parameters of PMPS to compute $\tau_{DM}$, we could have also used the parameters of The Pulsar Arecibo L-band Feed Array (PALFA) Survey, however the differences in the results in the end are not large if we use these parameters, consequently, it persists as a favorable approximation to exclusively utilize the parameters delineated by PMPS.

\subsubsection{$\gamma$-ray detection model}

The $\gamma$-ray emission model relies on the striped wind model, describing $\gamma$-ray photons production emanating from the current sheet within the striped wind. In order to detect the $\gamma$-ray, the line of sight of the observer must remain around the equator plane with an inclination angle constrained by $|\xi - \pi/2| \leq \alpha$.
The $\gamma$-ray luminosity is extracted from a study of \citet{Kalapotharakosetal}, where they showed that the luminosity is described by a fundamental plane. The 3D model depends on the magnetic field $B$, the spin down luminosity $\dot{E}$ and the cut-off energy $\epsilon_{cut}$. However, in our PPS the cut-off energy is not computed therefore we use their two dimensional version
\begin{equation} \label{eq:gamma_lum}
L_{\gamma(2D)} = 10^{26.15 \pm 2.6} \ W \ \left(\frac{B}{10^8 \ \text{T}}\right)^{0.11\pm 0.05} \ \left(\frac{\dot{E}}{10^{26} \ \text{W}}\right)^{0.51 \pm 0.09}
\end{equation}
The spindown $\dot{E}$ is computed with equation~(\ref{eq:spin_down_eq}) and the associated $\gamma$-ray flux detected on earth is computed with 
\begin{equation} \label{eq:flux_gamma}
F_{\gamma} = \frac{L_{\gamma(2D)}}{4\,\pi \, f_{\Omega}\, d^2}
\end{equation}
where $f_{\Omega}$ is a factor depending on the emission model reflecting the anisotropy. For the striped wind model \citet{Petri2011} showed that this factor varies between 0.22 and 1.90. Nevertheless, an approximation is made : if $\alpha < - \xi + 0.6109$ then $f_{\Omega}=1.9$ otherwise $f_{\Omega} = 1$, since as can be observed in Fig. 7 of \citet{Petri2011} $f_{\Omega}$ is usually equal to one of this two values depending on the value of the inclination angle $\alpha$, hence this approximation. The pulsar is detected in gamma depending on the instrumental sensitivity as described below. 

The sensitivity in $\gamma$-ray is based on the expectation of the Fermi/LAT instrument \footnote{\href{https://fermi.gsfc.nasa.gov/ssc/data/analysis/documentation/Cicerone/Cicerone_LAT_IRFs/LAT_sensitivity.html}{https://fermi.gsfc.nasa.gov/ssc/data/analysis/documentation/Cicerone \\ /Cicerone\_LAT\_IRFs/LAT\_sensitivity.html}}. Two conditions are required to count the detection, the first condition is that the source must be bright enough with a sufficiently high flux. If the galactic latitudes of the pulsar is < 2°, then $F_{\rm min} = 4 \times 10^{-15}$ W.m$^{-2}$, and if blind searches are assumed we set $F_{\rm min} = 16 \times 10^{-15}$ W.m$^{-2}$. Hence, if the $\gamma$-ray flux $F_{\gamma}$ is greater than $F_{\rm min}$, the first condition is fulfilled. The second condition concerns the dispersion measure. We set a threshold on the dispersion measure, DM (a similar approach was done in \citet{Gonthieretal} for millisecond $\gamma$-ray pulsars), because $\gamma$-ray canonical pulsars typically originate from supernova explosions and are associated with regions of recent star formation, such as supernova remnants or star-forming regions. These environments can have higher densities of free electrons, leading to higher DM values for pulsars in these regions, that is why we use this quantity as a discriminator for detection. Therefore, the DM threshold chosen is 15 cm$^{-3}.$pc, following the lowest value which can be found in the ATNF catalogue for high-energy pulsars. Thus, with these two conditions fulfilled, a pulsar is reported detected in $\gamma$-ray.

\section{Results} \label{S4} 

Let us now move to the PPS simulation results. First we show runs simulating 10~million pulsars with and without a death line, allowing to generate old pulsars with ages up to $4.1\times10^{8}$~yr. Then we show runs simulating only 1~million pulsars with ages up to $4.1\times10^7$~yr, also with and without a death line. We explore a possible spin-velocity misalignment effect due to the galactic potential and the impact of the initial spin period distribution, comparing normal vs log-normal distributions. Finally, the $\gamma$-ray simulations and detection are exposed in details.

\subsection{Simulating 10 millions pulsars} \label{sim_10M}

Let us start with a sample containing 10~millions pulsars showing the impact of the death line. A same study will be done for 1~million pulsars.

\subsubsection{Without a death line}

The first set of simulations discarded the death line. The parameters used for these runs are given in Table~\ref{tabl_sim_para}. The total number of pulsars is summarized in Table~\ref{tabl_nbpulsar_without_deathline}. 
\begin{table}[h]
\caption{Parameters used in the simulations.}
\label{tabl_sim_para} 
\centering 
\begin{tabular}{c c c c c c c} 
\hline\hline 
$\tau_{\text{birth}}$ (1/yr) & $P_{\text{mean}}$ (ms) & $B_{\text{mean}}$ (T) & $\sigma_p$ & $\sigma_b$ & $\alpha_d$ \\
\hline 
41 & 129 & $2.75 \times 10^8$ & 0.45 & 0.5 & 1.5 \\ 
\hline 
\end{tabular}
\end{table}
\begin{table}[h]
\caption{Number of pulsars detected without the implementation of the death line.}
\label{tabl_nbpulsar_without_deathline} 
\centering 
\begin{tabular}{c c c c c} 
\hline\hline 
log($\dot{E}$) (W) & $N_{tot}$ & $N_r$ & $N_g$ & $N_{rg}$\\
\hline 
>31 & 7 & 0 & 4 & 3\\ 
>28 & 161 & 4 & 87 & 63\\
Total & 2491 & 2169 & 163 & 159\\
\hline 
\end{tabular}
\end{table}
The quantities $N_r$, $N_g$, and $N_{rg}$ are the number of radio-only, $\gamma$-only, and radio-loud $\gamma$-ray pulsars, respectively,  extracted from our simulation.

Compared to the results obtained by \citet{Dirson22}, at very high spin down luminosity, $\dot{E} > 10^{31}$~W, the detection rate is almost identical but at $\dot{E} > 10^{28}$~W much less pulsars are detected in radio and radio/$\gamma$-ray in our work. There are probably two causes for this discrepancy in the number of detected pulsars at high spin down luminosities $\dot{E} > 10^{28}$~W : firstly the difference in the spin period distribution at birth. Pulsars have a higher probability to start with a higher $P_0$ since the mean of the distribution is higher (in the previous study the mean period was 60~ms while we set it to 129~ms), therefore the slowing down of the pulsars can only decrease their rotation frequency $\Omega$, and equation~\eqref{eq:spin_down_eq} shows that $\dot{E}$ depends on~$\Omega$. However, it does not affect the detection rate in $\gamma$-rays because usually canonical $\gamma$-ray pulsars possess a high $\dot{P}$ and a low $P$, being young, therefore their $\dot{E}$ is higher than for radio pulsars. Secondly, the ISM dispersion and scattering reduces the number of detected radio pulsars, results without the ISM effect are not shown but discarding its impact would lead to much more detected radio pulsars. Nonetheless, in terms of total number of detected radio pulsars, including the Galactic potential increases this number compared to \citet{Dirson22}, a result expected because of the attractive nature of the potential of the Galaxy. Moreover in this latter version, the PPS moved pulsars in random spatial directions at constant speed, as a result no pulsar was bound to the Milky Way in closed orbits an artifact that is especially true for old pulsars which could have time to leave the Galaxy.

\begin{figure}[h]
\includegraphics[width=\columnwidth]{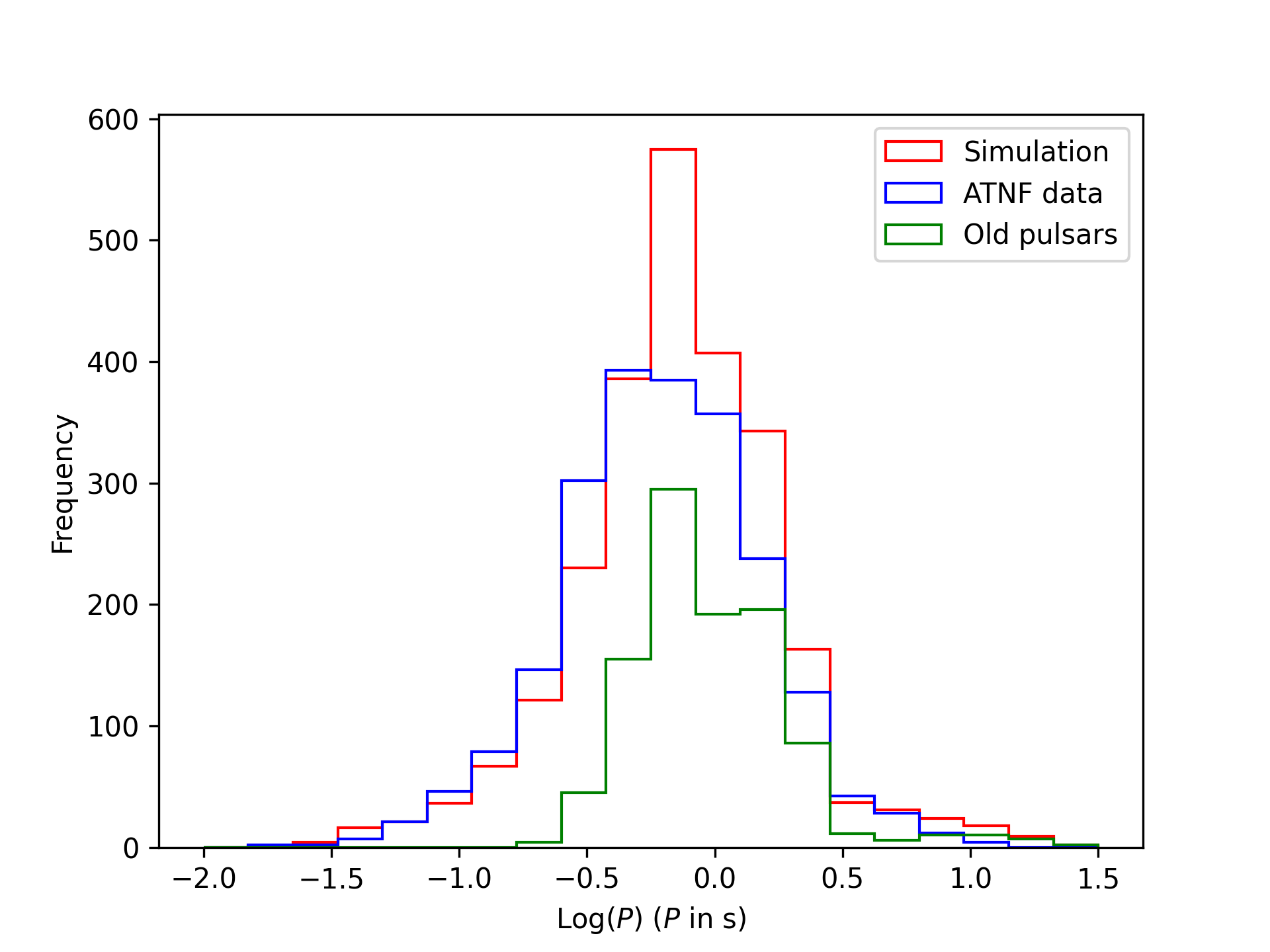}
\caption{Distribution of the observed period taken from the ATNF catalogue,
along with the simulations without the implementation of the death line. In green, the period of the simulated pulsars with characteristic age greater than $10^{8}$~yr.}
\label{histo_period_nodeath}
\end{figure}

\begin{figure}[h]
\includegraphics[width=\columnwidth]{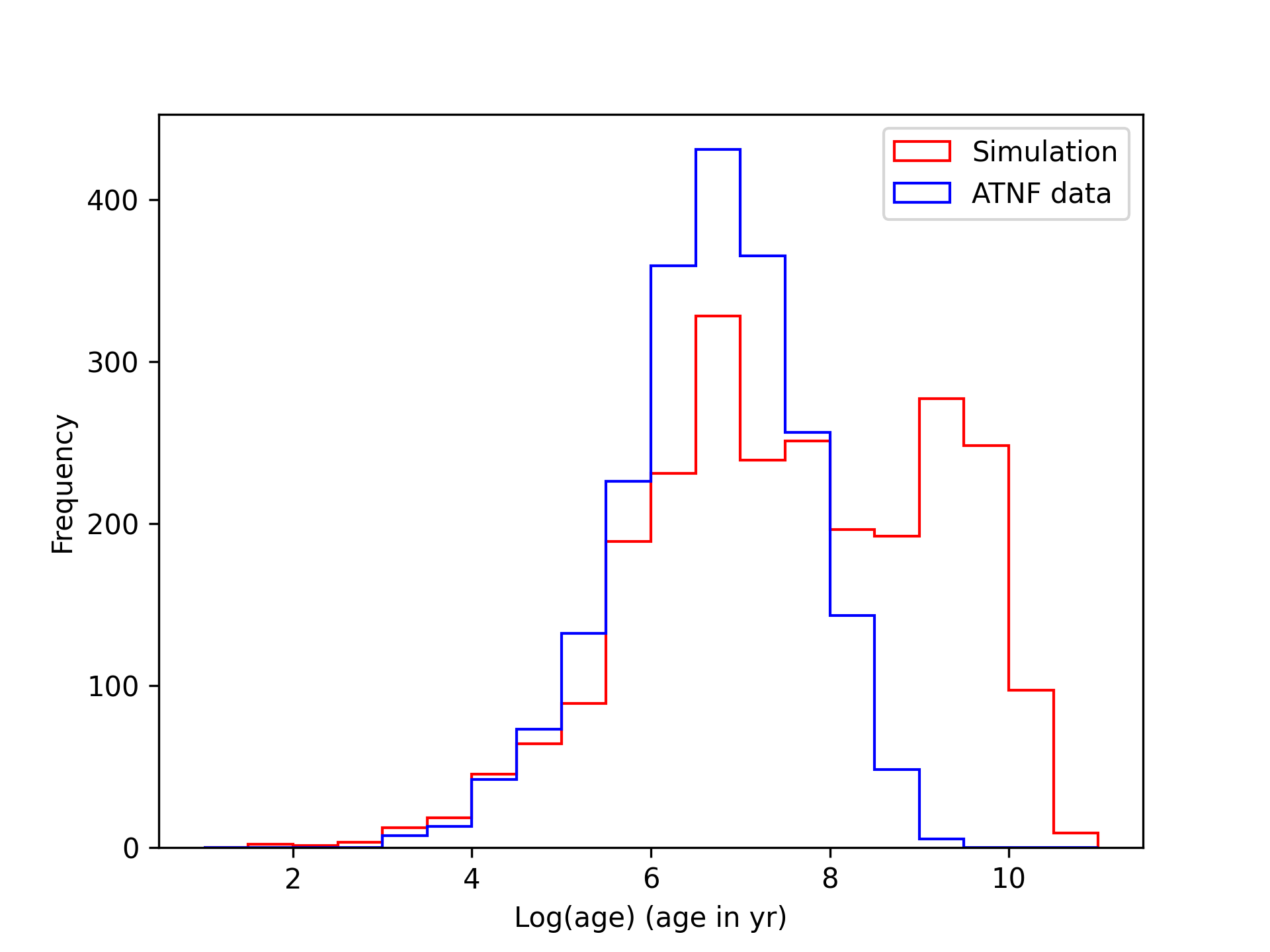}
\caption{Distribution of the observed age taken from the ATNF catalogue,
along with the simulations without the implementation of the death line.}
\label{histo_age_nodeath}
\end{figure}

Fig.~\ref{histo_period_nodeath} compares the distribution of simulated and observed spin periods, in red and blue respectively. However the comparison is not satisfactory because of an excess of pulsars detected in the simulation. Furthermore, this excess is also present in Fig.~\ref{histo_age_nodeath} where the characteristic age of the observed pulsars and the simulated pulsars are plotted. Note that we employ the characteristic ages defined by
\begin{equation} \label{charac_age}
\tau_c = \frac{P}{2 \dot{P}}
\end{equation}
because we usually do not know the real ages of the observed pulsars.

\begin{figure}[h]
\resizebox{\hsize}{!}{\includegraphics{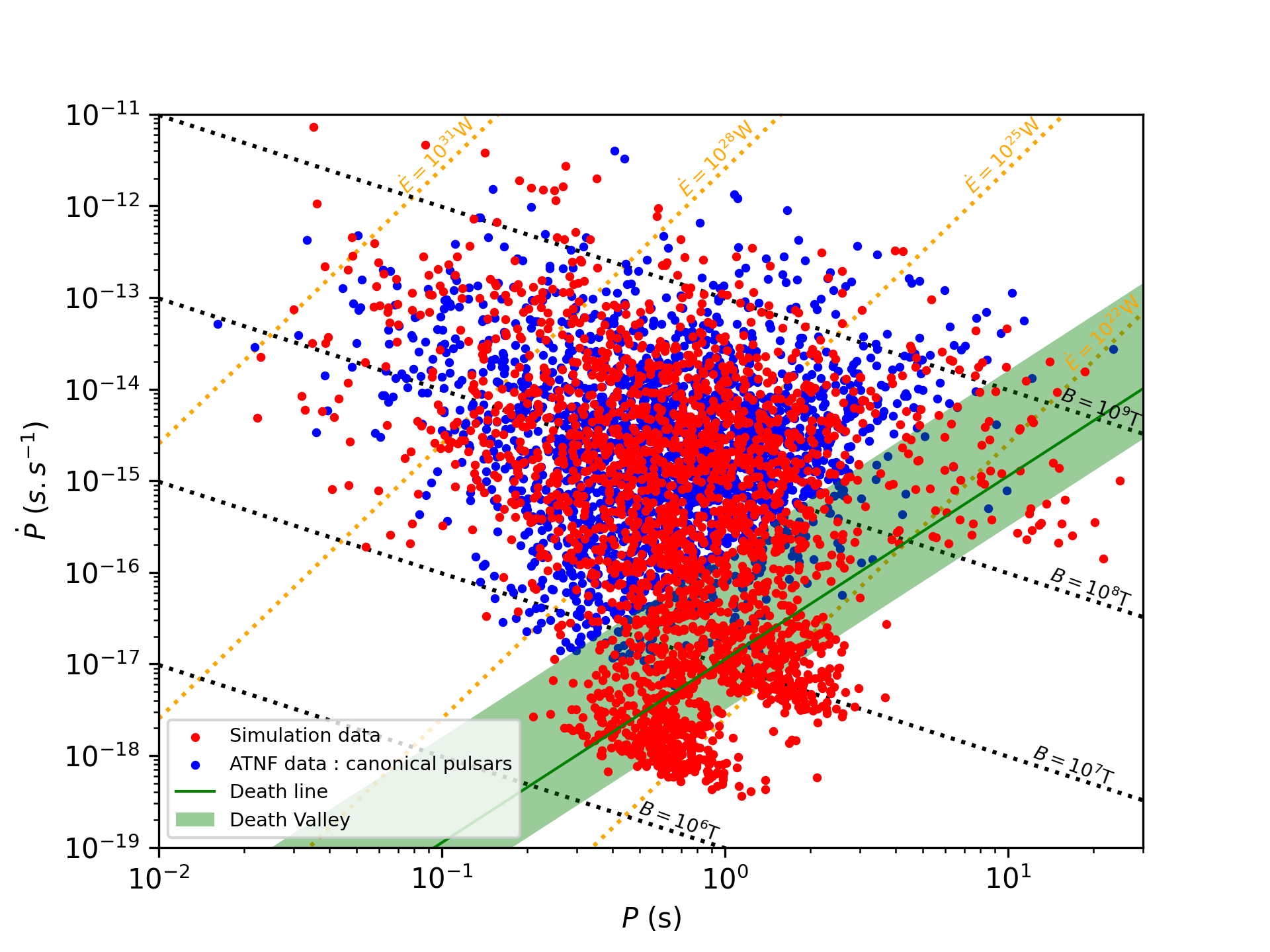}}
\caption{$P-\dot{P}$ diagram of the simulated population, along with the observations, without the implementation of the death line. \\ \vspace{4mm}}
\label{ppdot_no_death}
\end{figure}

\begin{figure}[h]
\resizebox{\hsize}{!}{\includegraphics{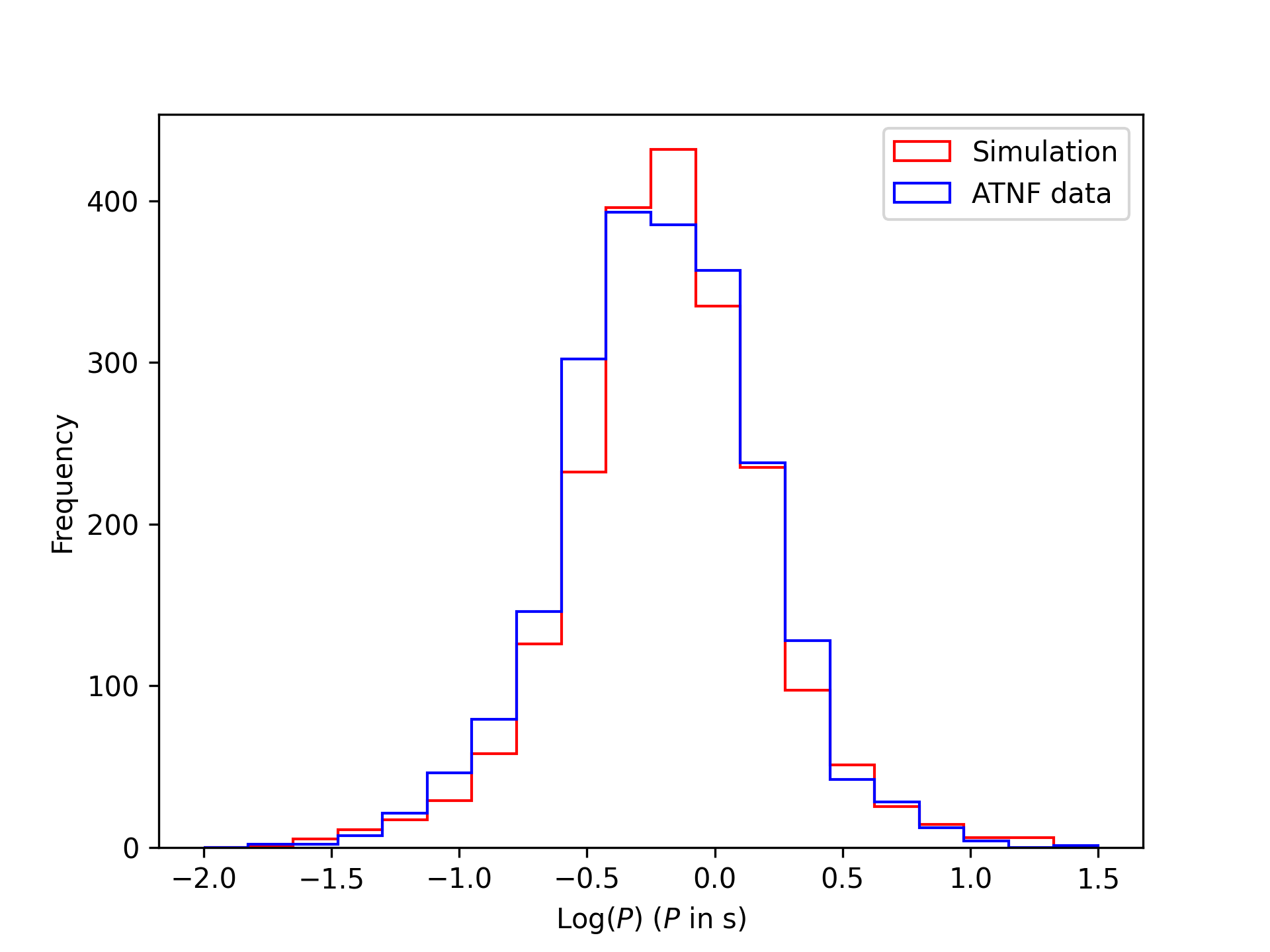}}
\caption{Distribution of the observed period taken from the ATNF catalogue,
along with the simulations with the death line implementation.}
\label{histo_period_death}
\end{figure}

In order to verify whether there is a correlation between old pulsars and those with a high spin period or not, Fig.~\ref{histo_period_nodeath} shows in green pulsars with a characteristic age greater than $10^{8}$~yr, corresponding to the oldest pulsars of the simulation. It coincides with the excess found in Fig.~\ref{histo_period_nodeath}. We conclude that canonical pulsars older than $10^8$~yr measured as their real age in our simulations should be discarded at the end of the simulation, since the characteristic age is overestimating the real age of pulsars. \modif{Indeed, it appears that we are unable to detect canonical pulsars older than $10^8$~yr.} 
The Galactic potential forbids many pulsars to escape from the Milky Way, they remain bound to the Galaxy, in contrast to the older pulsars found in \citet{Dirson22}, that were escaping the Galaxy easily because of the absence of this potential. Fig.~\ref{ppdot_no_death} shows that many pulsars are in what is commonly called the graveyard of pulsars at the bottom right of the $P-\dot{P}$ diagram, the death valley being shown in Fig.~\ref{ppdot_no_death} but not implemented.
Given this excess of long period pulsars when neglecting the death line, we decided to implement it, in order to stick closer to reality.

\subsubsection{With the death line} 

\begin{figure}[h]
\resizebox{\hsize}{!}{\includegraphics{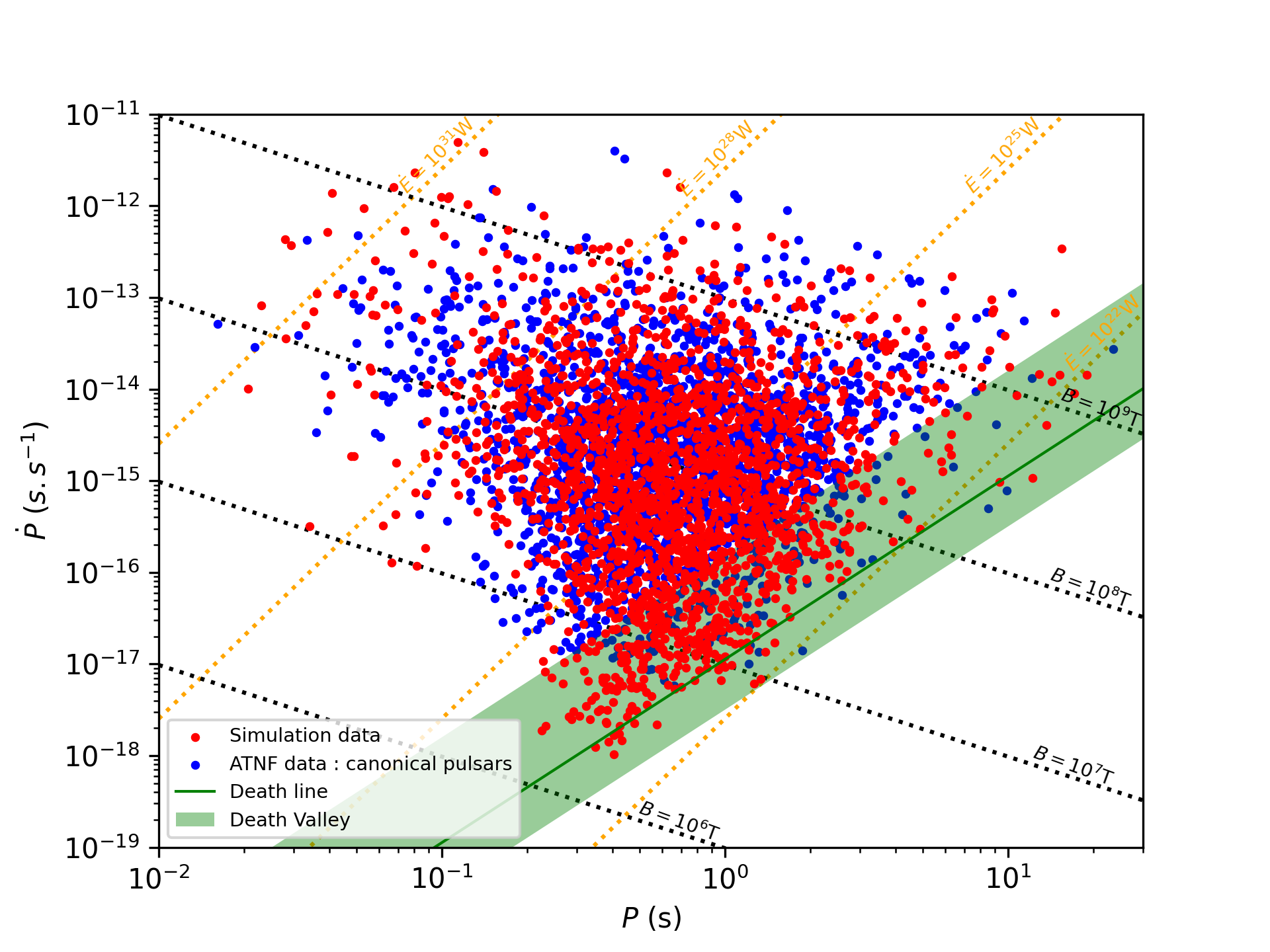}}
\caption{$P-\dot{P}$ diagram of the simulated population, along with the observations, with the death line implementation.}
\label{ppdot_death}
\end{figure}

Adding a death line decreases the number of pulsars detected in radio compared to the situation without a death line, see Table~\ref{tabl_nbpulsar_with_deathline}.
The death line improves the fit as seen in Fig.~\ref{histo_period_death} for the distribution of spin period and compared with Fig.~\ref{histo_period_nodeath}, even though there is a slight excess of pulsars which have spin period between $10^{-0.5}$~s and $1$~s. In Fig.~\ref{ppdot_death}, these pulsars in excess lie at $\dot{P}$ between $10^{-18}$ and $10^{-17}$, where there is no data and which correspond to the oldest pulsars in the simulation. Therefore the oldest pulsars simulated here do not correspond to any real data. In addition, too many pulsars end up close to the death line, disagreeing with the observations. The death line creates a pile up effect not seen in the observations.
\begin{table}[h]
\caption{Number of pulsars detected with the death line implementation.}
\label{tabl_nbpulsar_with_deathline} 
\centering 
\begin{tabular}{c c c c c} 
\hline\hline 
log($\dot{E}$) (W) & $N_{tot}$ & $N_r$ & $N_g$ & $N_{rg}$\\
\hline 
>31 & 10 & 0 & 5 & 5\\ 
>28 & 118 & 7 & 69 & 42\\
Total & 2077 & 1807 & 116 & 154\\
\hline 
\end{tabular}
\end{table}
\begin{figure}[h]
\resizebox{\hsize}{!}{\includegraphics{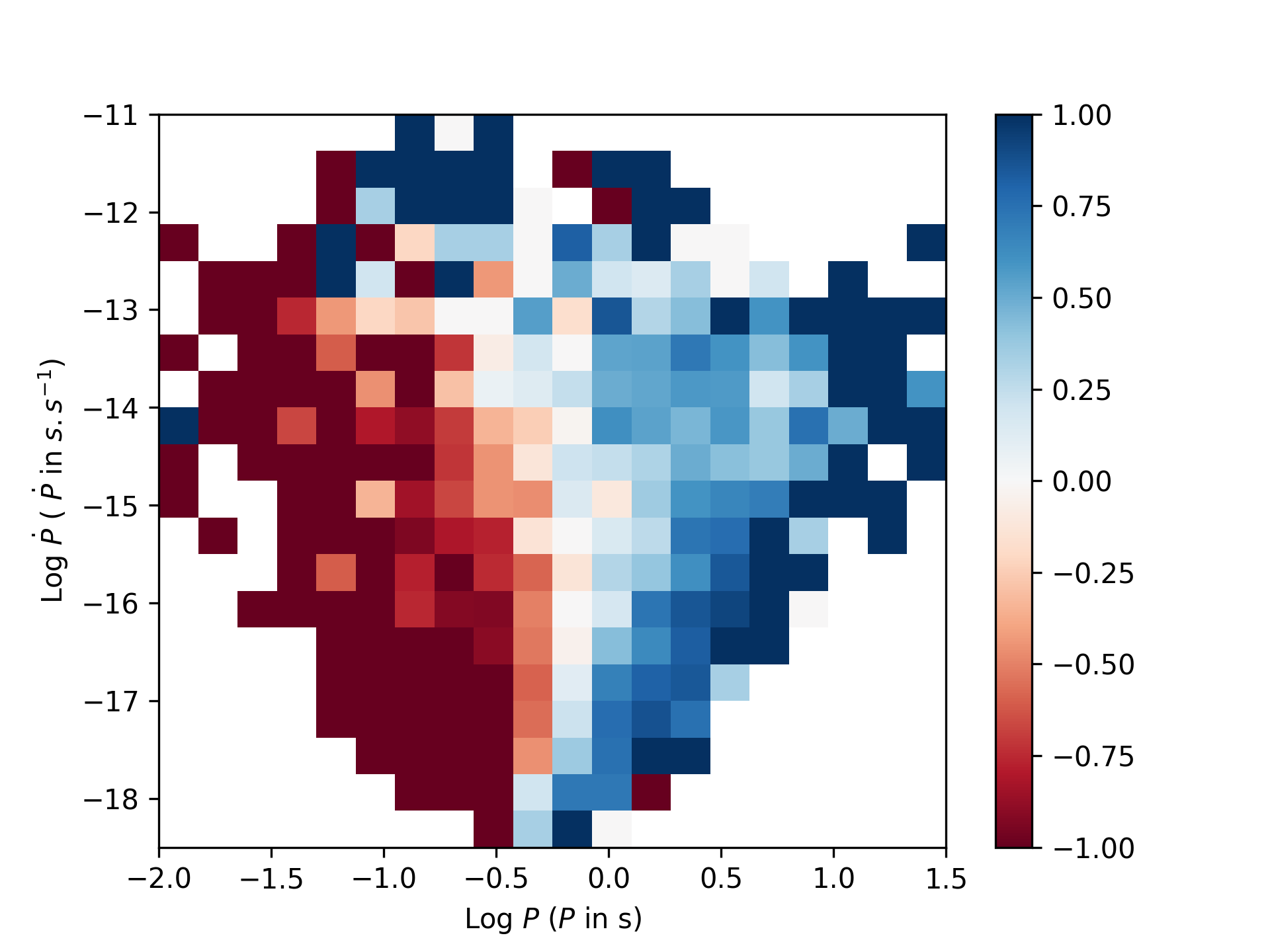}}
\caption{Density plot of the $P-\dot{P}$ diagram in comparison with observations.}
\label{density_plot_PPdot}
\end{figure}
The goodness-of-fit is checked against a density plot of the $P-\dot{P}$ diagram as in \citet{Johnston2017} instead of putting points for individual pulsars. In order to construct such diagrams we binned the data evenly in a log scale in $P$ and $\dot{P}$ and count the number of pulsars in each bin. This leads to a 2D histogram to be compared with the observed density plot 2D histogram. Because the total number of simulated pulsars $N_{totsim}$ is different from the total number of observed pulsars $N_{totobs}$, we normalise the number of pulsars in each bin in order to compare proportions. We therefore introduce the quantity $R$ as follows
\begin{equation} \label{eq:R_value}
R = \frac{N_{sim} / N_{totsim} - N_{obs}/ N_{totobs}}{ N_{sim} / N_{totsim} + N_{obs} / N_{totobs}} .
\end{equation}
With this definition the number $R$ remains in the interval $[-1,+1]$.
$N_{sim}$ is the number of pulsar count in a bin of the simulation, $N_{obs}$ the number of pulsar detected in a bin of the observations. 
A good fit in a bin corresponds to a value of $R$ close to $0$, meaning that the simulation results are close to the observations. Inspecting fig.~\ref{density_plot_PPdot}, we note a stronger deviation from the observations at the boundaries of the density plot, confirming for instance the pile up close to the death line.
The high values of $R$ at these boundaries originates from the weak number of pulsars counted in these bins leading to bad statistics and large variations in the proportions, a side effect of the density plot approach. The fact that the colors are very dark, shows that the proportion of pulsars in most of the bins are not good. 

A Kolmogorov Smirnov (KS) test \citep{Smirnov} was conducted on both $P$ and $\dot{P}$ which gave p-values substantially below 0.05 for both quantities, when no death line is implemented. While we obtained p-values of 0.05 for $P$ and well below 0.05 for $\dot{P}$ with a death line implementation. The null hypothesis that observations and simulations originate from the same distribution for $P$ can not be rejected (p-value $\geq0.05$) only when a death line is considered, but is clearly rejected for $\dot{P}$ in both case. 

\subsection{Simulating 1 millions pulsars} \label{sim_1M}
\subsubsection{With the death line}

The parameters used for these runs are shown in Table~\ref{tabl_sim_para}. This new set of simulations is intended to verify if generating pulsars younger than $4.1\times10^{7}$~yr is sufficient, since older pulsars are barely observed. The death line is also taken into account. Without any specifications, in all figures, with the exception of fig.~\ref{histo_dist_death_1M}, \ref{positions_pulsars_death_1M} and \ref{histo_latitude_death_1M}, simulations were performed while considering the ISM effect.
\begin{table}[h]
\caption{Number of pulsars detected with the death line implementation for 1 million pulsars simulated.}
\label{tabl_nbpulsar_with_deathline_1M} 
\centering 
\begin{tabular}{c c c c c} 
\hline\hline 
log($\dot{E}$) (W) & $N_{tot}$ & $N_r$ & $N_g$ & $N_{rg}$\\
\hline 
>31 & 8 & 0 & 7 & 1\\ 
>28 & 138 & 6 & 82 & 52\\
Total & 1645 & 1338 & 165 & 142\\
\hline 
\end{tabular}
\end{table}

\begin{figure}[h]
\resizebox{\hsize}{!}{\includegraphics{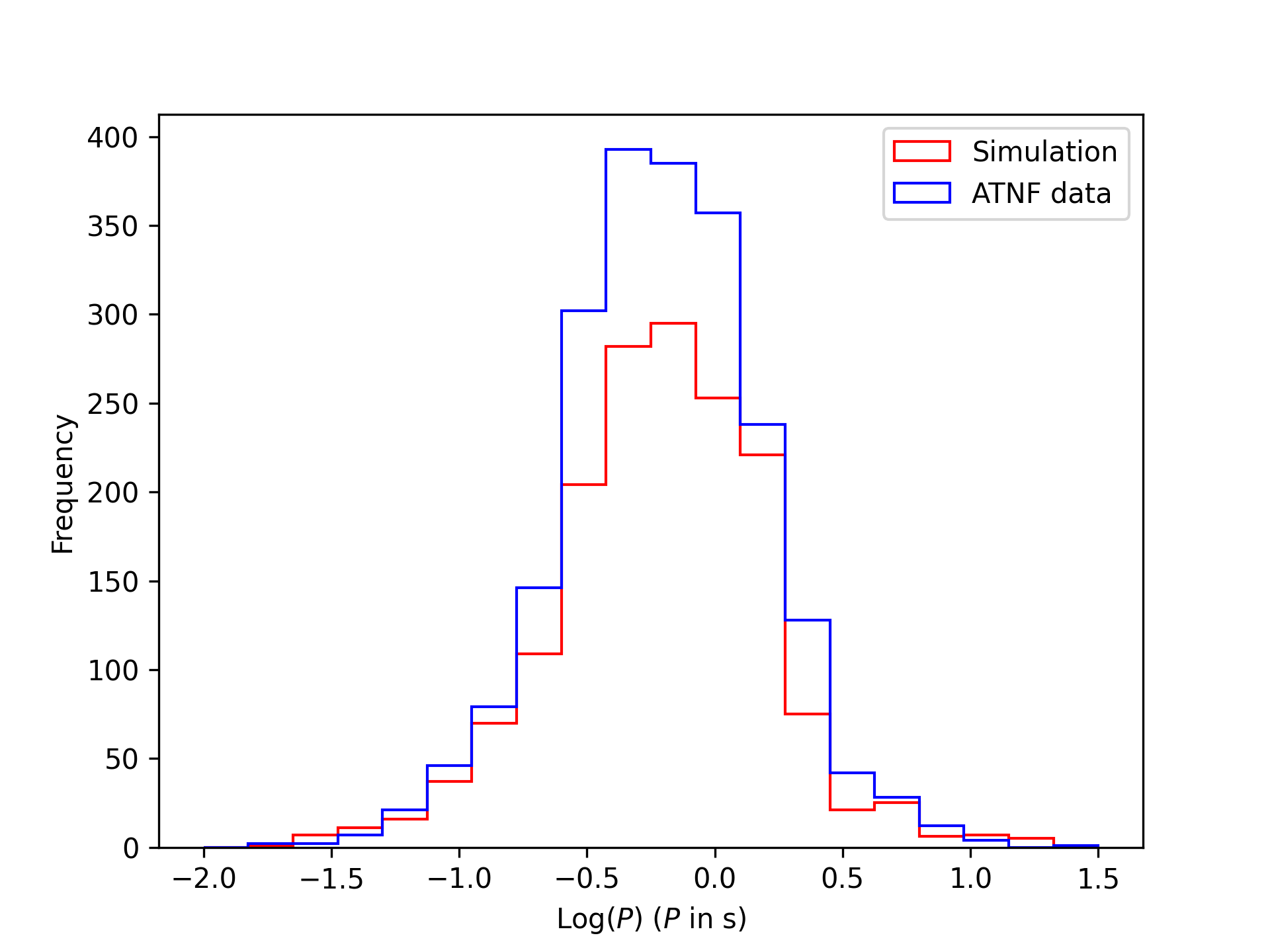}}
\caption{Distribution of the observed period taken from the ATNF catalogue,
along with the simulations with the death line implementation for 1 million pulsars simulated.}
\label{histo_P_death_1M}
\end{figure}

\begin{figure}[h]
\resizebox{\hsize}{!}{\includegraphics{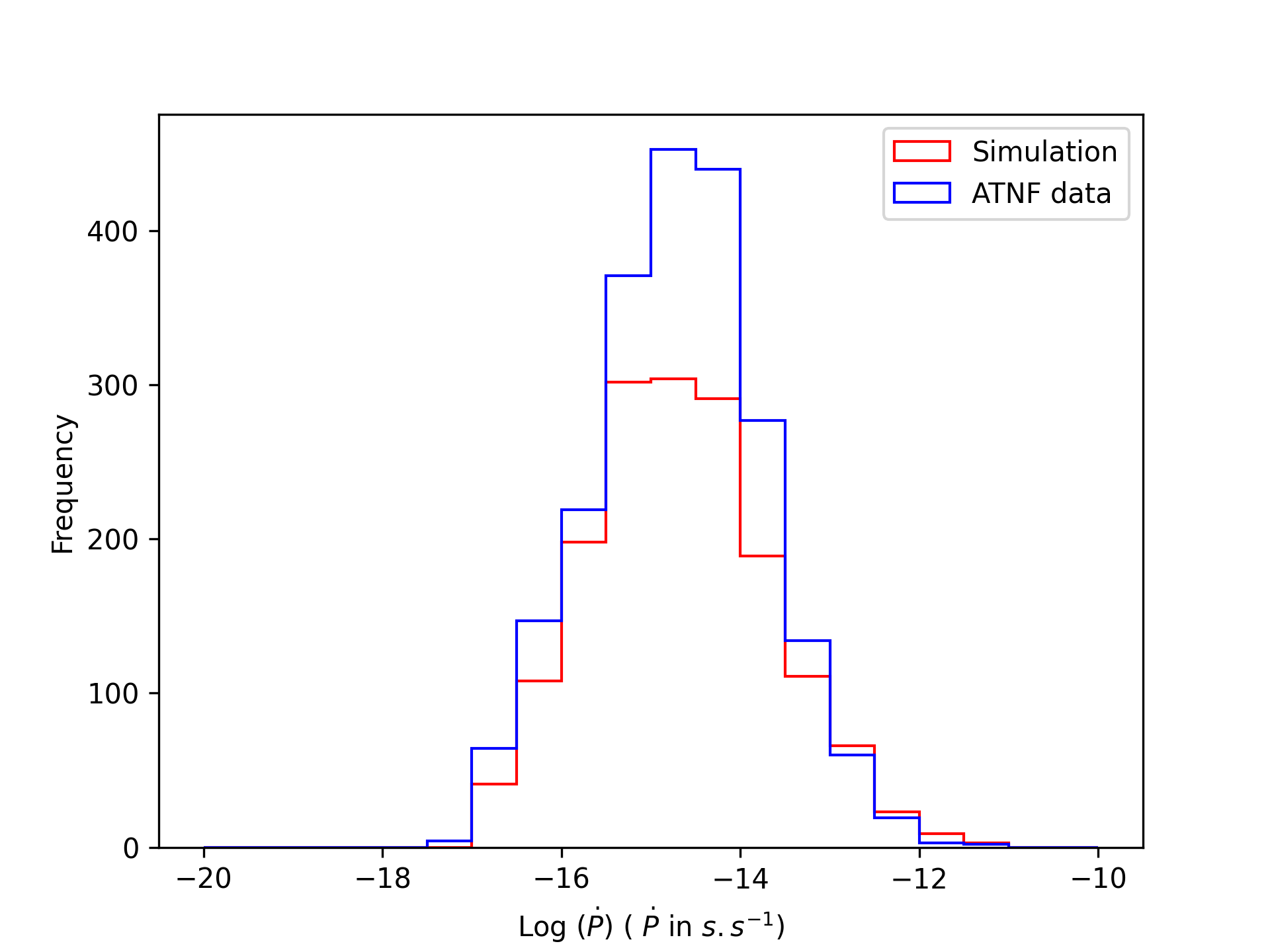}}
\caption{Distribution of the observed period derivative taken from the ATNF catalogue,
along with the simulations with the death line implementation for 1 million pulsars simulated.}
\label{histo_Pdot_death_1M}
\end{figure}

\begin{figure}[h]
\resizebox{\hsize}{!}{\includegraphics{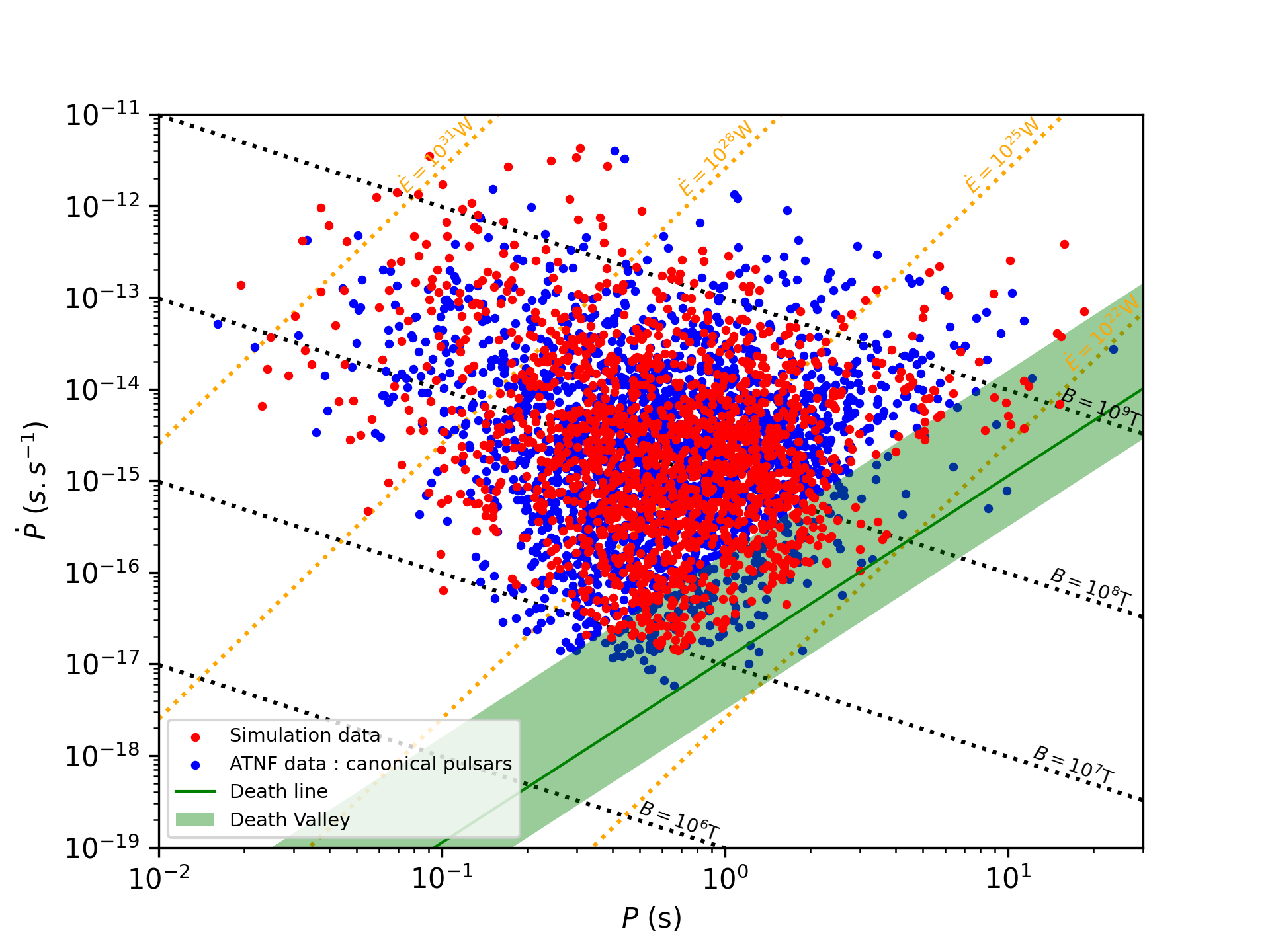}}
\caption{$P-\dot{P}$ diagram of the simulated population, along with the observations, with the death line implementation for 1 million pulsars simulated.}
\label{histo_PPdot_death_1M}
\end{figure}

Table~\ref{tabl_nbpulsar_with_deathline_1M} shows that a few less pulsars were detected in this simulation compared to the one with the death line and 10 millions pulsars (see Table \ref{tabl_nbpulsar_with_deathline}). Moreover, inspecting the histograms for spin period and spin period derivative respectively in Fig. \ref{histo_P_death_1M} and \ref{histo_Pdot_death_1M}, we conclude that observations and simulations are very similar in both histograms. A KS test was conducted for both distributions $P$ and $\dot{P}$ and yielded p-values of $0.67$ and $0.17$ respectively. Therefore, it is impossible to reject the null hypothesis stating that the simulated distributions are similar to the observed distribution.

\begin{figure}[h]
\resizebox{\hsize}{!}{\includegraphics{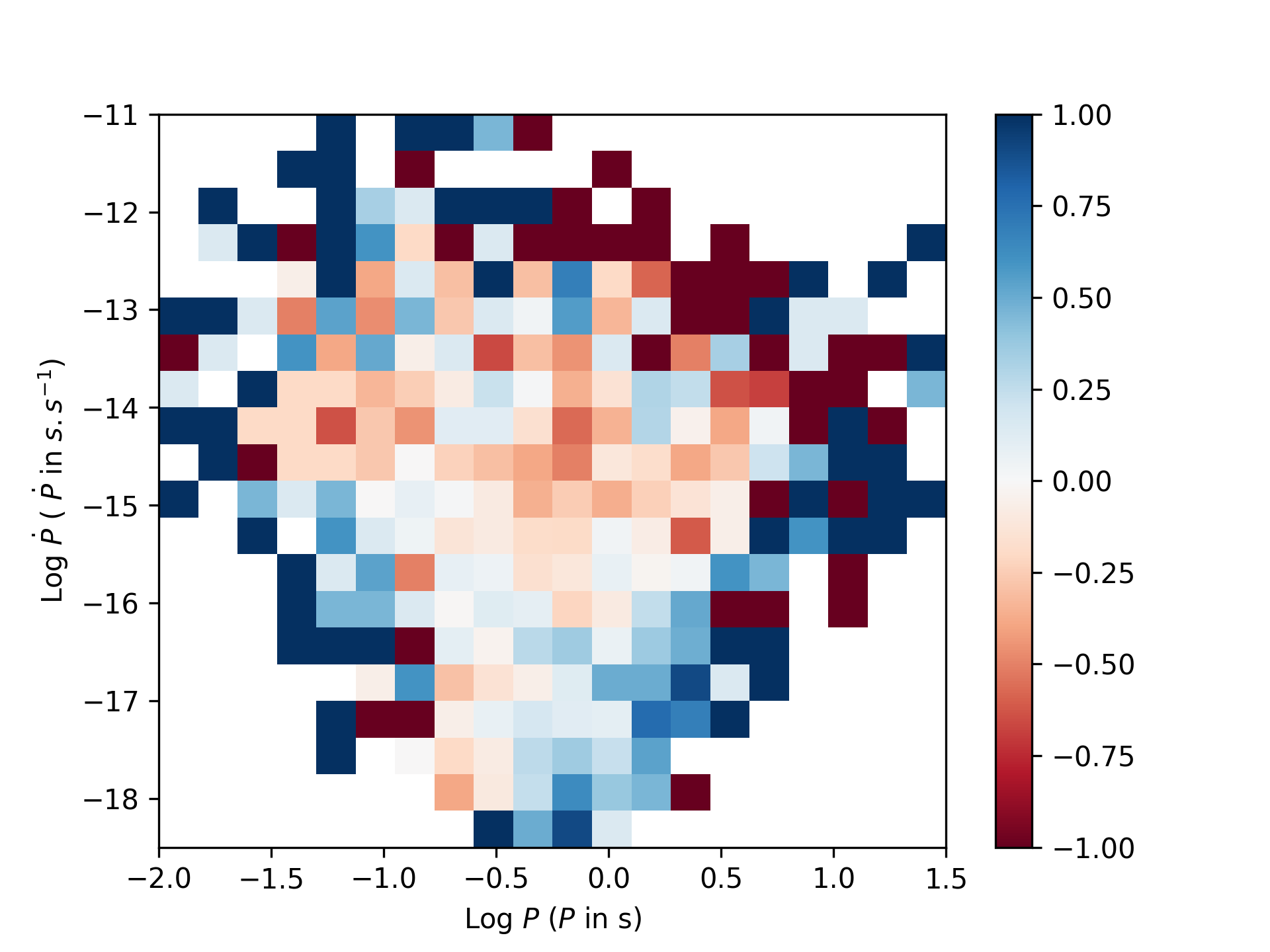}}
\caption{Density plot of the $P-\dot{P}$ diagram in comparison with observations for 1 million pulsars simulated. \\}
\label{density_plot_PPdot_1M}
\end{figure}

Fig~\ref{histo_PPdot_death_1M} shows the good agreement between simulations and observations.
Furthermore, comparing Fig.~\ref{density_plot_PPdot_1M} and Fig.~\ref{density_plot_PPdot}, the former performs better because globally the values of $R$ are close to zero, highlighting the fact that the proportion of pulsars in each 2D bin are almost alike. In addition, the pile up effect close to the death line has been removed from the $P-\dot{P}$ diagram, which better fits to the data.

\begin{figure}[h]
\resizebox{\hsize}{!}{\includegraphics{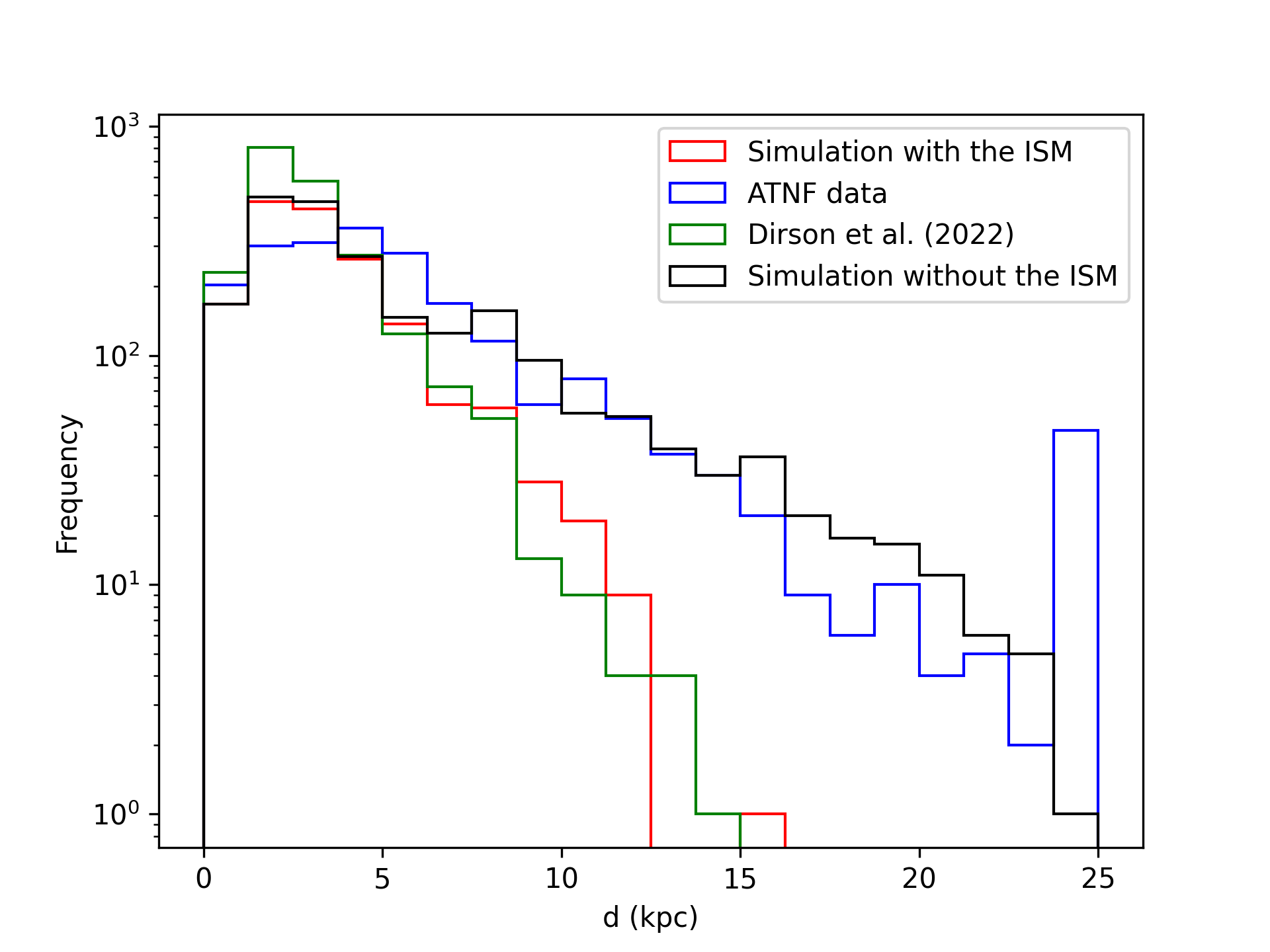}}
\caption{Distance distribution to Earth for observed and simulated populations (1 million pulsars), including the death line implementation. \\}
\label{histo_dist_death_1M}
\end{figure}

The pulsar distance distribution to Earth is shown in Fig.~\ref{histo_dist_death_1M}. The improvement compared to \cite{Dirson22} is marginal. Anyway, let us remind that the estimated distances of the observed pulsars are subject to large uncertainties because of the distances derived from the dispersion measure values (DM) are highly uncertain, while the distances for the simulated pulsars are correct by construction. As a consequence, the comparison between the observed and simulated distance distribution is not a reliable indicator to check the correctness of our model as the discrepancies are subject to observational uncertainties. Fortunately the $P-\dot{P}$ improved compared to the previous work. This could mean the observed pulsars are closer than expected. Indeed, Fig.~\ref{positions_pulsars_death_1M} shows that the positions of the simulated pulsars in the X-Y plane of the Galaxy, including the ISM effect. They are closer to the Sun than the observed pulsars. Surprisingly, by removing the effect of dispersion and scattering of the ISM on the radio pulse profile, we obtain very similar distance distributions between observations and simulations, see Fig.~\ref{histo_dist_death_1M}. Therefore, a refinement of the ISM impact on radio pulse profiles might help getting a more realistic model.
  
\begin{figure}[h]
\includegraphics[width=\columnwidth]{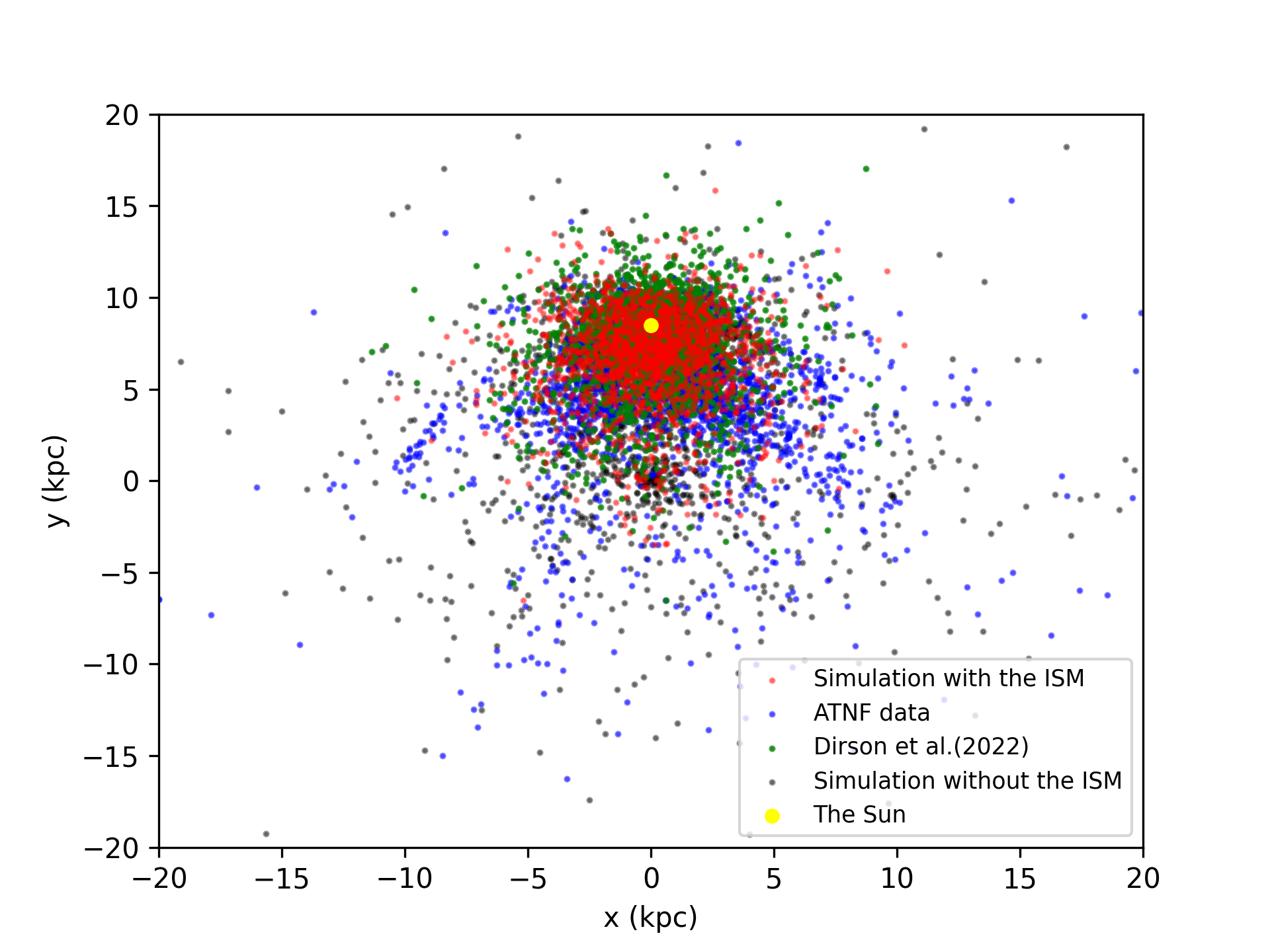}
\caption{Spatial distribution of the simulated pulsars, along with the observations, projected onto the Galactic plane for 1 million pulsars simulated.}
\label{positions_pulsars_death_1M}
\end{figure} 
\begin{figure}[h]
\resizebox{\hsize}{!}{\includegraphics{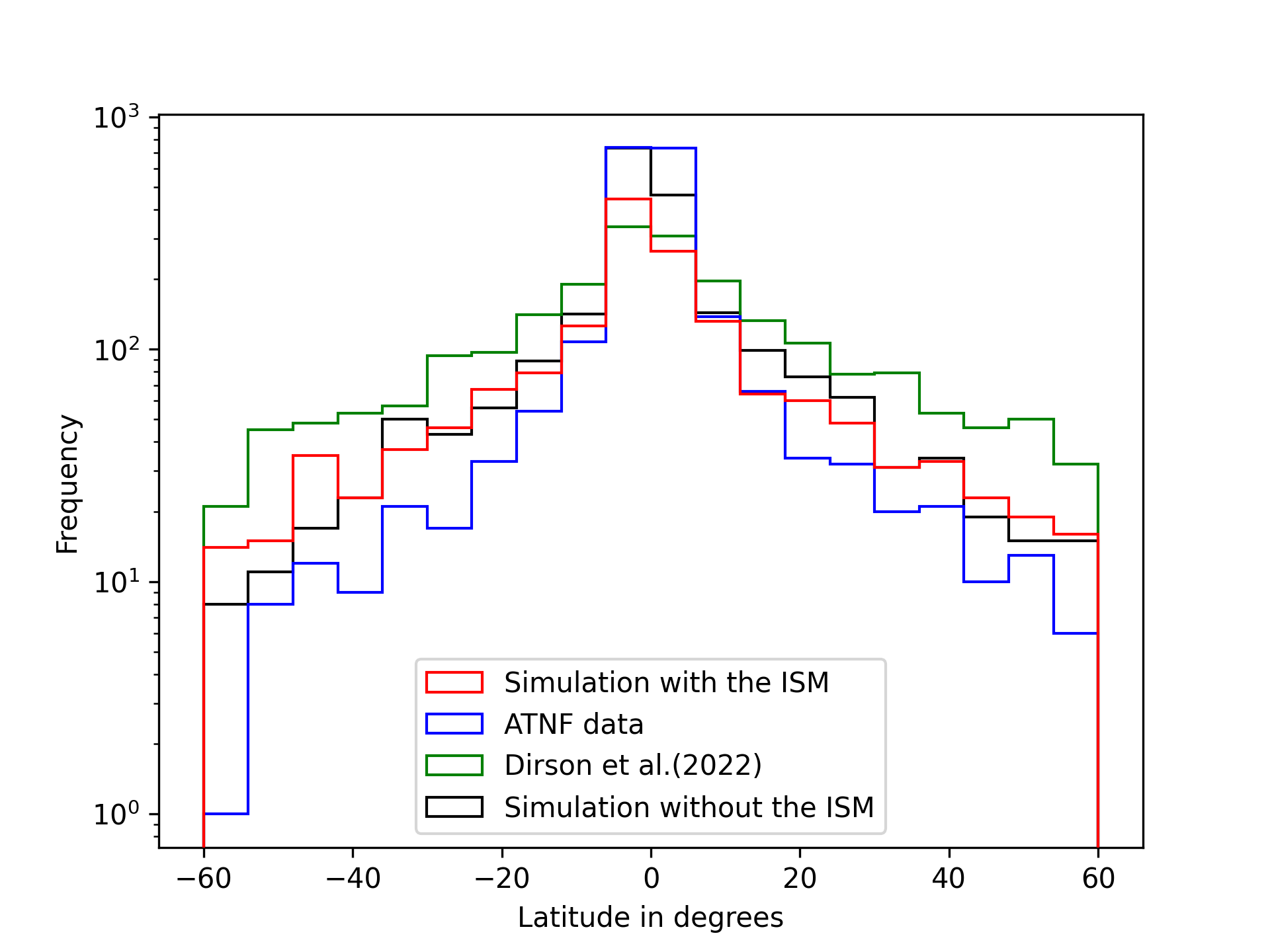}}
\caption{Distribution of the observed latitude taken from the ATNF catalogue,
along with the simulations with the death line implementation for 1 million pulsars simulated.}
\label{histo_latitude_death_1M}
\end{figure}  

Finally fig.~\ref{histo_latitude_death_1M} compares the Galactic latitude distributions with the study of \citet{Dirson22}, their Fig.~9. In our work the simulated distribution better matches the observations, especially around all the latitudes away from 0°. Therefore, the Galactic potential helped tracking the pulsar trajectories more realistically. We notice also the ISM removes pulsars close to a latitude of 0°, because the electron density is highest in the galactic plane of the Milky Way.

\subsubsection{Without a death line}

\begin{figure}[h]
\resizebox{\hsize}{!}{\includegraphics{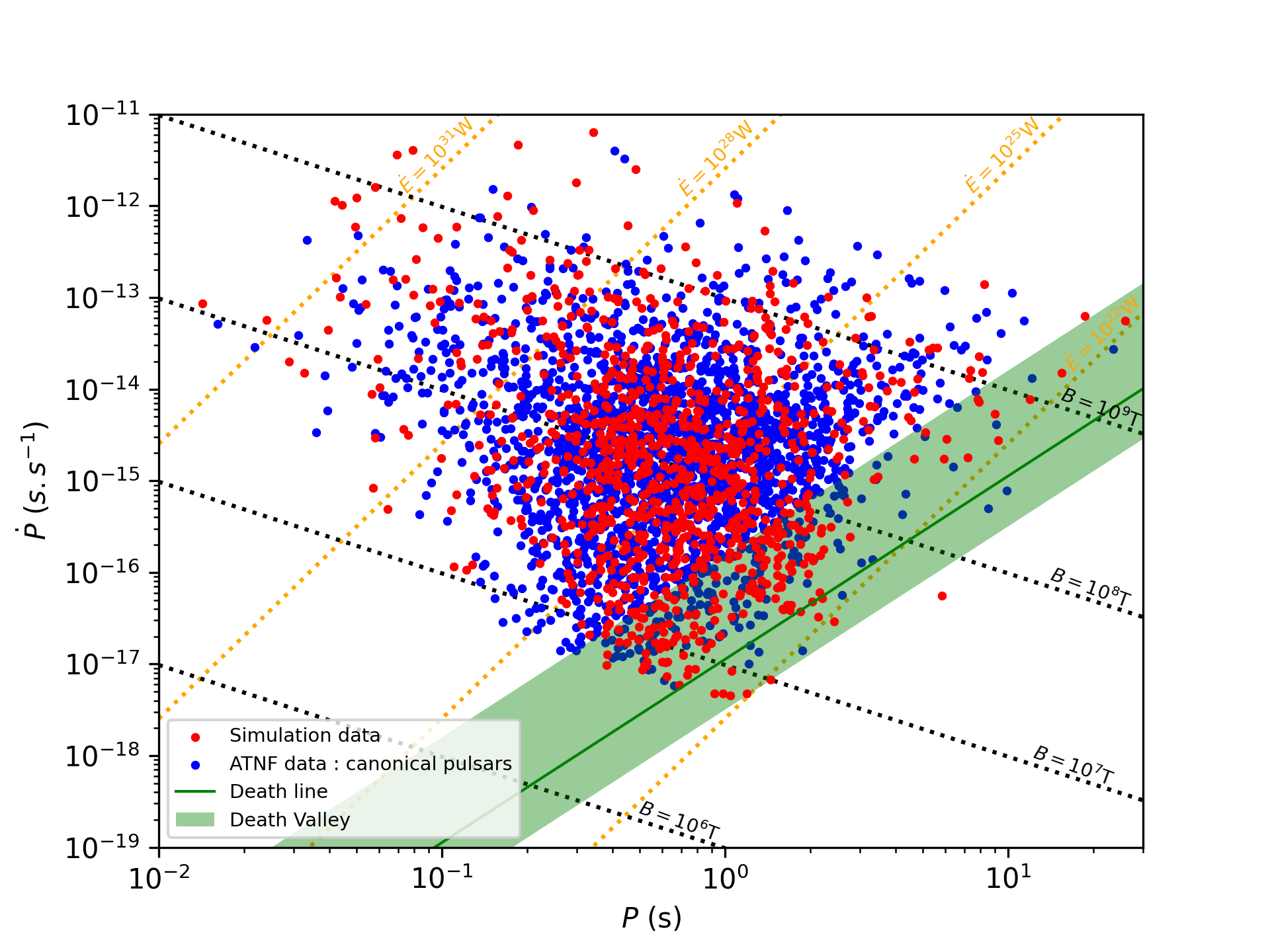}}
\caption{$P-\dot{P}$ diagram of the simulated population, along with the observations, without death line implementation, birth rate of 1/70 yr$^{-1}$ for 1 million pulsars simulated.}
\label{PPdot_1M_no_death_70br}
\end{figure}

\begin{figure}[h]
\resizebox{\hsize}{!}{\includegraphics{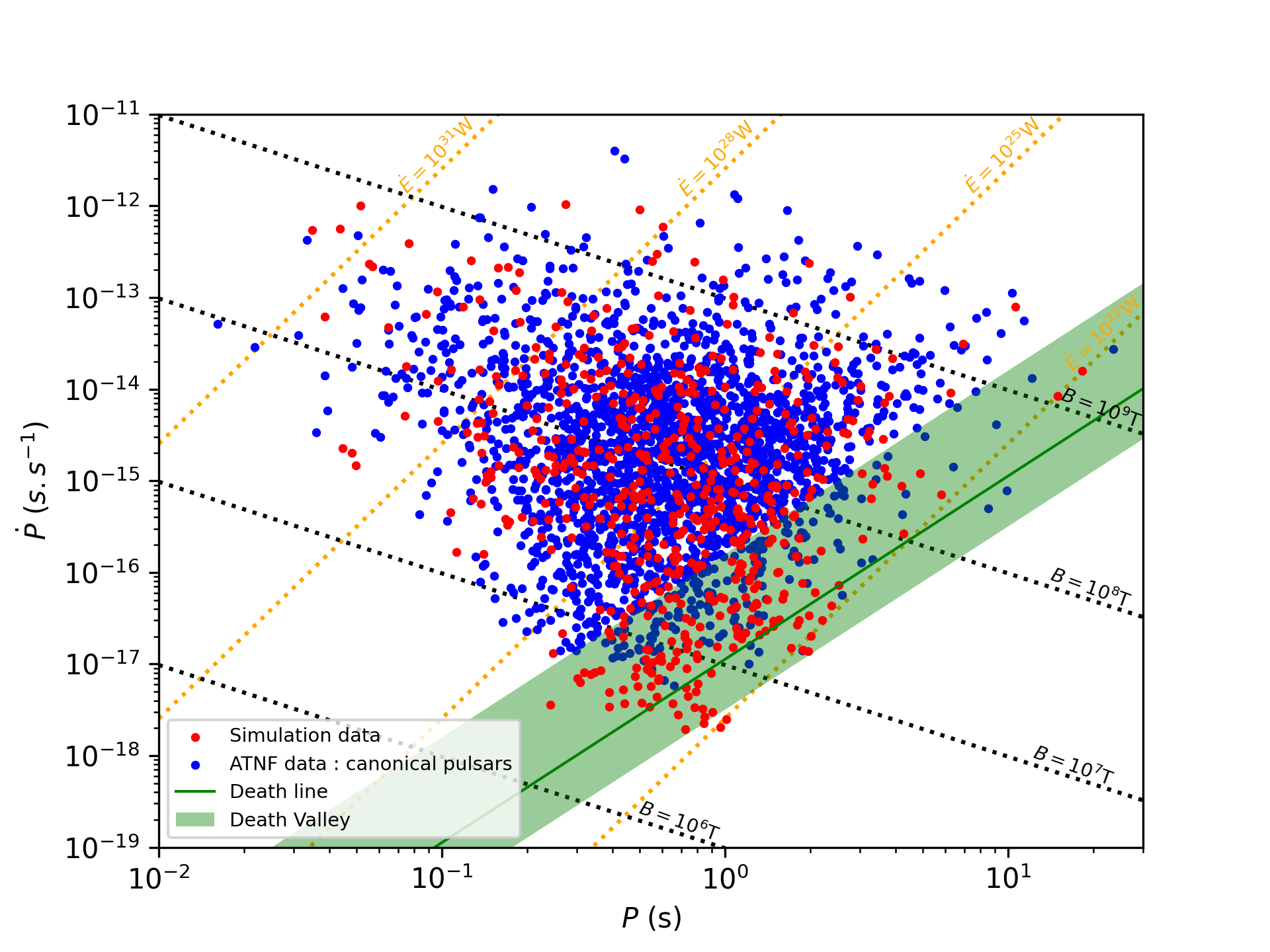}}
\caption{$P-\dot{P}$ diagram of the simulated population, along with the observations, without death line implementation, birth rate of 1/150 yr$^{-1}$ for 1 million pulsars simulated.}
\label{PPdot_1M_no_death_150br}
\end{figure}

Simulation with 1~million simulated pulsars does not allow to generate pulsars older than $4.1\times10^7$~yr. We therefore undertook a thorough examination of the usefulness of a death line in modeling the canonical pulsar population. Upon conducting simulations with and without a death line (see Table \ref{tabl_sim_para} for the parameter values used in the simulations), we found that the resultant number of detected pulsars and the corresponding plots exhibited remarkable similarities. By conducting the KS test on $P$ and $\dot{P}$ between two simulations, one with and one without the death line, we obtained p-values of 0.20 and 0.16, respectively. Since both p-values are greater than 0.05, this indicates no significant difference between the two simulations, supporting their similarity. Consequently, our attention turned towards exploring the impact of lower birth rates to ascertain if they provide a more congruous fit for both spin period and spin period derivative distributions.

Running simulations with a birth rate of $1/70$~yr$^{-1}$ or a birth rate of $1/150$~yr$^{-1}$ (as a consequence, increasing the oldest simulated pulsar), conspicuously revealed an insufficient number of detected pulsars in both Fig.~\ref{PPdot_1M_no_death_70br} and Fig.~\ref{PPdot_1M_no_death_150br}. Intriguingly, a discernible paucity of pulsars congregating in the bottom right quadrant of the diagram, colloquially known as the pulsar graveyard, persisted even under these conditions, therefore regardless of the birth rates, thus casting doubt on the necessity of a death line in this configuration. Ultimately, our analysis underscores the pivotal importance of delineating the maximum real age of simulated pulsars, as elucidated in Subsection \ref{sim_10M}, while the incorporation of a death line do not seem indispensable for attaining realistic results when simulating 1 million pulsars.

\subsection{Spin-velocity misalignment effect} \label{spinvel_results}

With the aim of observing the effect of the Galactic potential, we looked at the alignment of the rotation axis and the proper motion velocities of the pulsars. Results are taken from the simulation containing 1~million pulsars, see Table \ref{tabl_sim_para} for the parameters used. \citet{Noutsosetal} showed by studying a catalogue of 58~pulsars that pulsars younger than 10~Myr keep their rotation axis and their velocities vector still aligned or anti-aligned, because they were aligned at birth according to \citet{Rankin}. Meanwhile, pulsars older than 10~Myr do not show this trend anymore. This progressive misalignment effect is due to the movement through the Galactic potential. 

\begin{figure}[h]
\resizebox{\hsize}{!}{\includegraphics{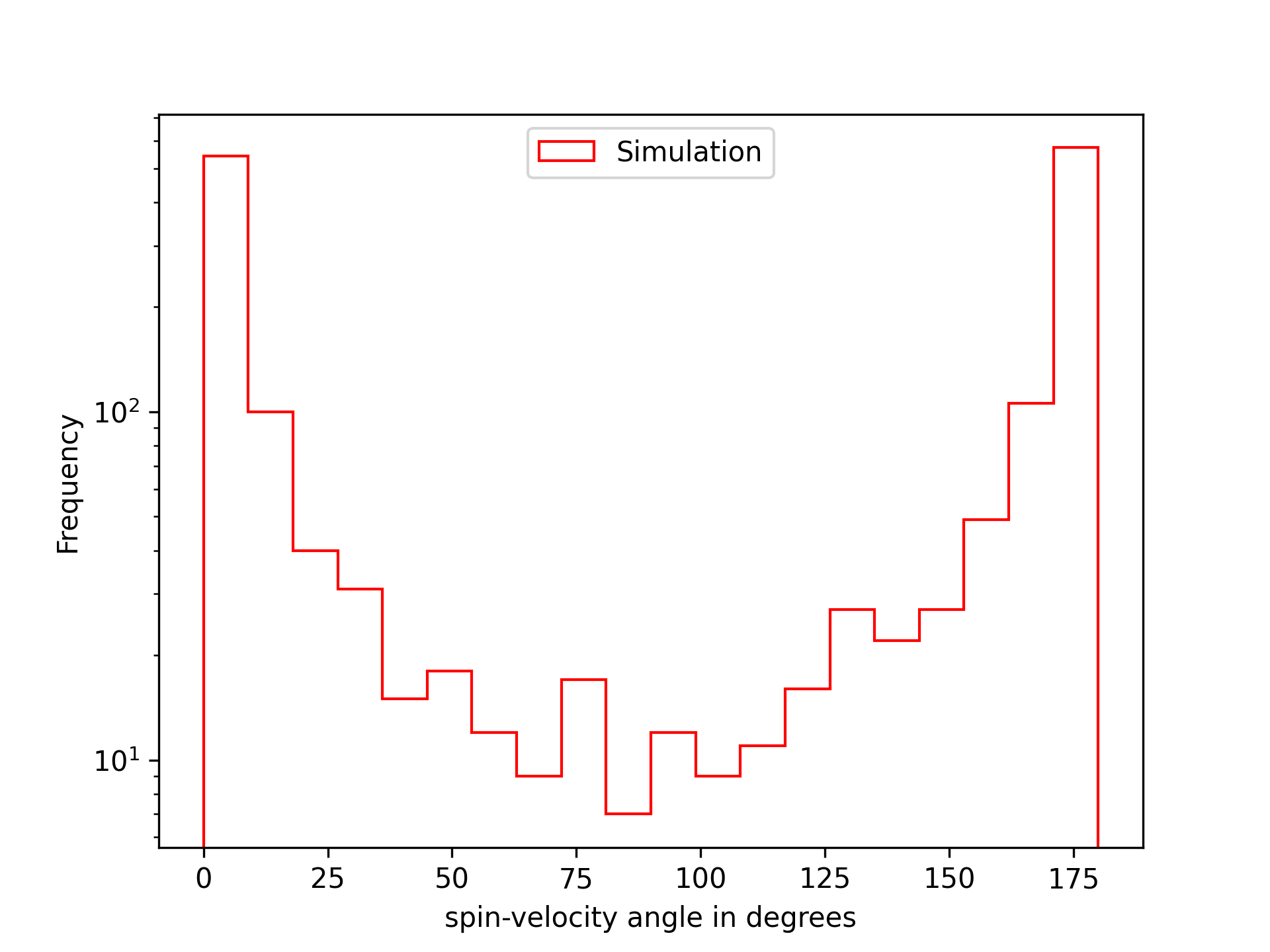}}
\caption{Distribution of the spin-velocity angles for all the detected pulsars in the simulation for 1 million pulsars simulated.}
\label{histo_spinvelangleall}
\end{figure}

\begin{figure}[h]
\resizebox{\hsize}{!}{\includegraphics{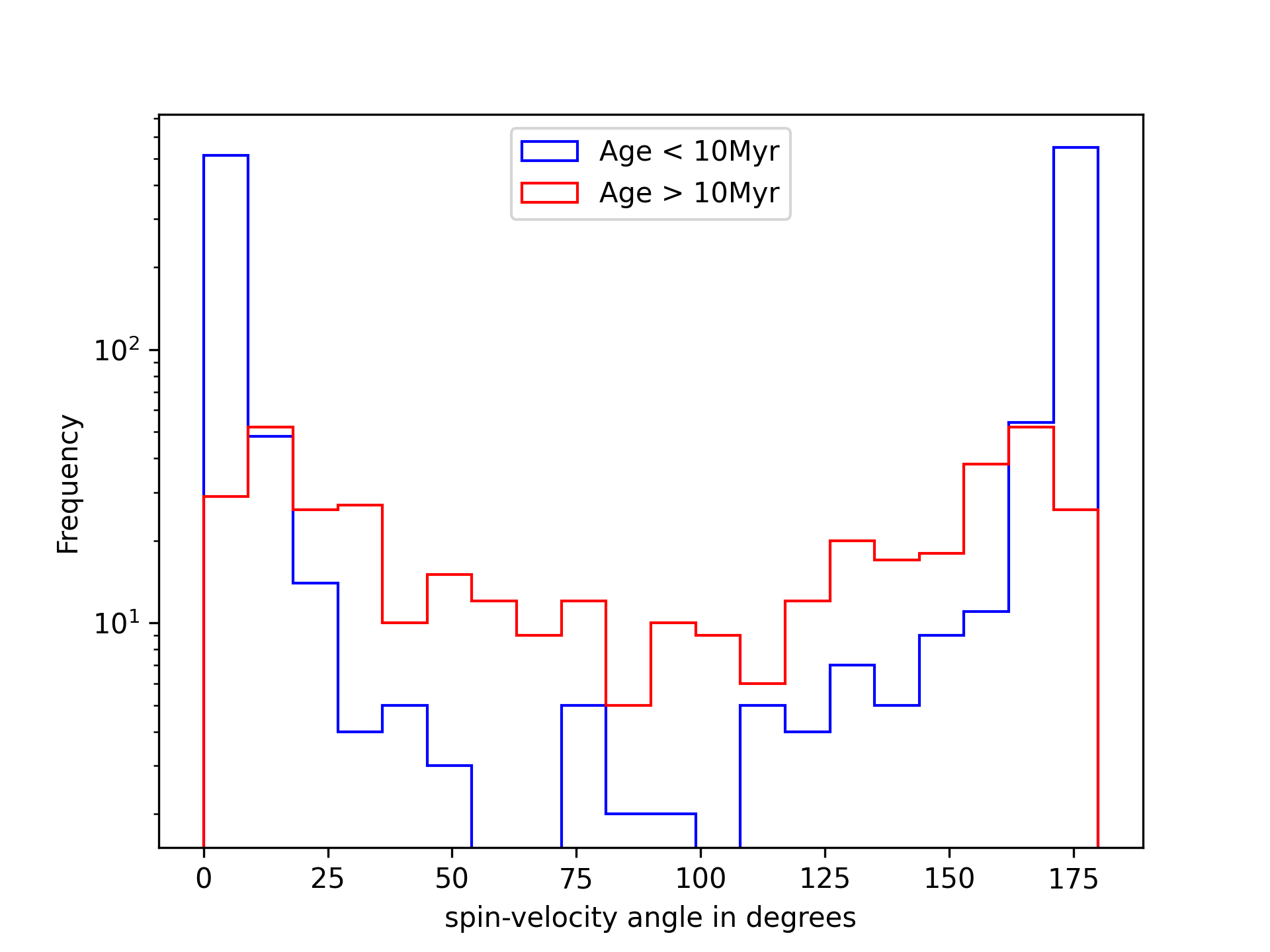}}
\caption{Distribution of the spin-velocity angles for the detected pulsars younger than 10 Myr (blue) and pulsars older than 10 Myr (red) in the simulation.}
\label{histo_spinvelangleyoungold}
\end{figure}

Fig.~\ref{histo_spinvelangleall} demonstrates that most pulsars have their rotation axis aligned or counter-aligned with their velocities, but for spin-velocity angles between 30° and 150° the numbers of pulsars is considerably lower than around alignment 0° and counter-alignment 180°. On the one hand, for pulsars younger than 10~Myr, shown in blue in Fig.~\ref{histo_spinvelangleyoungold}, only a small fraction of pulsars have their spin-velocity angle greater than 30° and below 150°. On the other hand, for pulsars older than 10~Myr, shown in red in Fig.~\ref{histo_spinvelangleyoungold}, their spin-velocity angle becomes greater than 30° or smaller than 150°.

\begin{figure}[h]
\resizebox{\hsize}{!}{\includegraphics{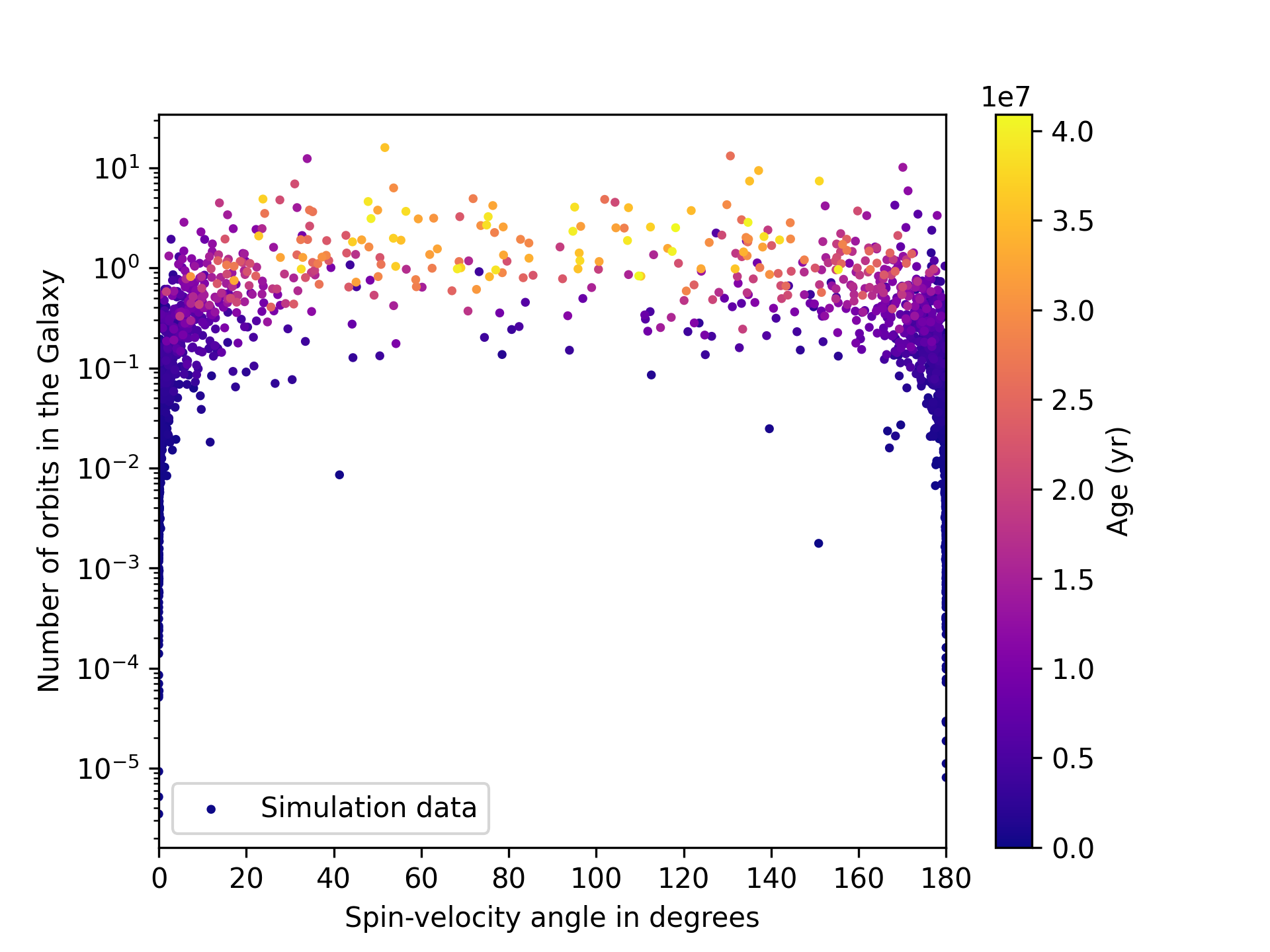}}
\caption{The spin-velocity angles for the detected pulsars in function of the number of orbits made by the pulsars.}
\label{spinvelorbitageplot}
\end{figure}

Furthermore, Fig.~\ref{spinvelorbitageplot} confirms that, the older the pulsar is, the better its chance to perform several orbits in the Galaxy, allowing it to significantly deviate from spin-velocity alignment or anti-alignment. In addition, even for less old pulsars able to perform many orbits in the Galaxy, their spin-velocity angle will evolve between 30° and 150°. This supports the idea that the Galactic potential is responsible for the misalignment irrespective of the age of the pulsar. The opposite is also true, some old pulsars do not perform many orbits (but most of them are), and as a consequence they keep approximately their spin-velocity alignment or misalignment. The previous plots clearly highlight the influence of the Galactic potential making pulsars lose their alignment between their rotation axis and their proper motion velocities through time. 

\subsection{Log normal vs normal distribution for spin period at birth}

\begin{table}[h]
\caption{P-values and total number of pulsars detected ($N_{detection}$) obtained for simulations with different $\bar{P}$ for the Gaussian initial distribution of spin period.}
\label{tabl_nbpulsar_with_deathline_gaussian_distribution} 
\centering 
\begin{tabular}{c c c c c} 
\hline\hline 
$\bar{P}$ (ms) & $\sigma_p$ (ms) & p-value (P) & p-value ($\dot{P}$)& $N_{detection}$ \\
\hline 
60 & 30 & $6.58\times10^{-20}$ & $3.76\times10^{-4}$ & 1512\\ 
80 & 30 & $1.84\times10^{-15}$ & 0.051 & 1671\\
100 & 30 & $2.34\times10^{-10}$ & 0.123 & 1721\\
129 & 30 & $1.02\times10^{-6}$ & 0.050 & 1787\\
140 & 30 & $1.83\times10^{-5}$ & 0.006 & 1816\\
\hline 
\end{tabular}
\end{table}
In order to estimate the impact of the spin period distribution at birth, we compare the results of the log-normal law with the previously used normal law in \citet{Dirson22}
\begin{equation} \label{eq:gauss_distrib}
    p(P_0) = \frac{1}{\sigma_p \sqrt{2\pi}} e^{-(P_0-\bar{P})^2/(2\sigma_p)^2} \ .
\end{equation}
The simulation in this subsection uses the parameters listed in the two first columns of Table~\ref{tabl_nbpulsar_with_deathline_gaussian_distribution} and to be put into equation~\eqref{eq:gauss_distrib}. The Galactic potential and the death line were both taken into account to compare the influence of the initial spin period distribution on the 1~million pulsars simulated, see Table \ref{tabl_sim_para} for the parameters used. In Table~\ref{tabl_nbpulsar_with_deathline_gaussian_distribution}, only results for $\sigma_p = 30$~ms are shown because other values have been tried out, namely 10~ms, 20~ms and 40 ms but gave no higher p-values than for $\sigma_p=30$~ms. Moreover, we chose $\bar{P}$ from 60~ms to 140~ms because it is the range usually used in other work \citep{FaucherG,Gullon14,Dirson22}.

Table~\ref{tabl_nbpulsar_with_deathline_gaussian_distribution} highlights the fact that for $\bar{P}$ above 60~ms and below 140~ms, an acceptable p-value $\geq 0.05$ is reached for the $\dot{P}$ distribution. However no matter the value of $\bar{P}$, a satisfying p-value for the $P$ distribution is not reached in contrary to the log-normal distribution at birth, for which satisfying p-value for both distributions are obtained. 
We conclude that the initial distribution for the spin period found by \citet{Igo} allows to better reproduce the canonical pulsars population. 

\subsection{$\gamma$-ray detection} \label{gamma_results}

\begin{figure}[h]
\resizebox{\hsize}{!}{\includegraphics{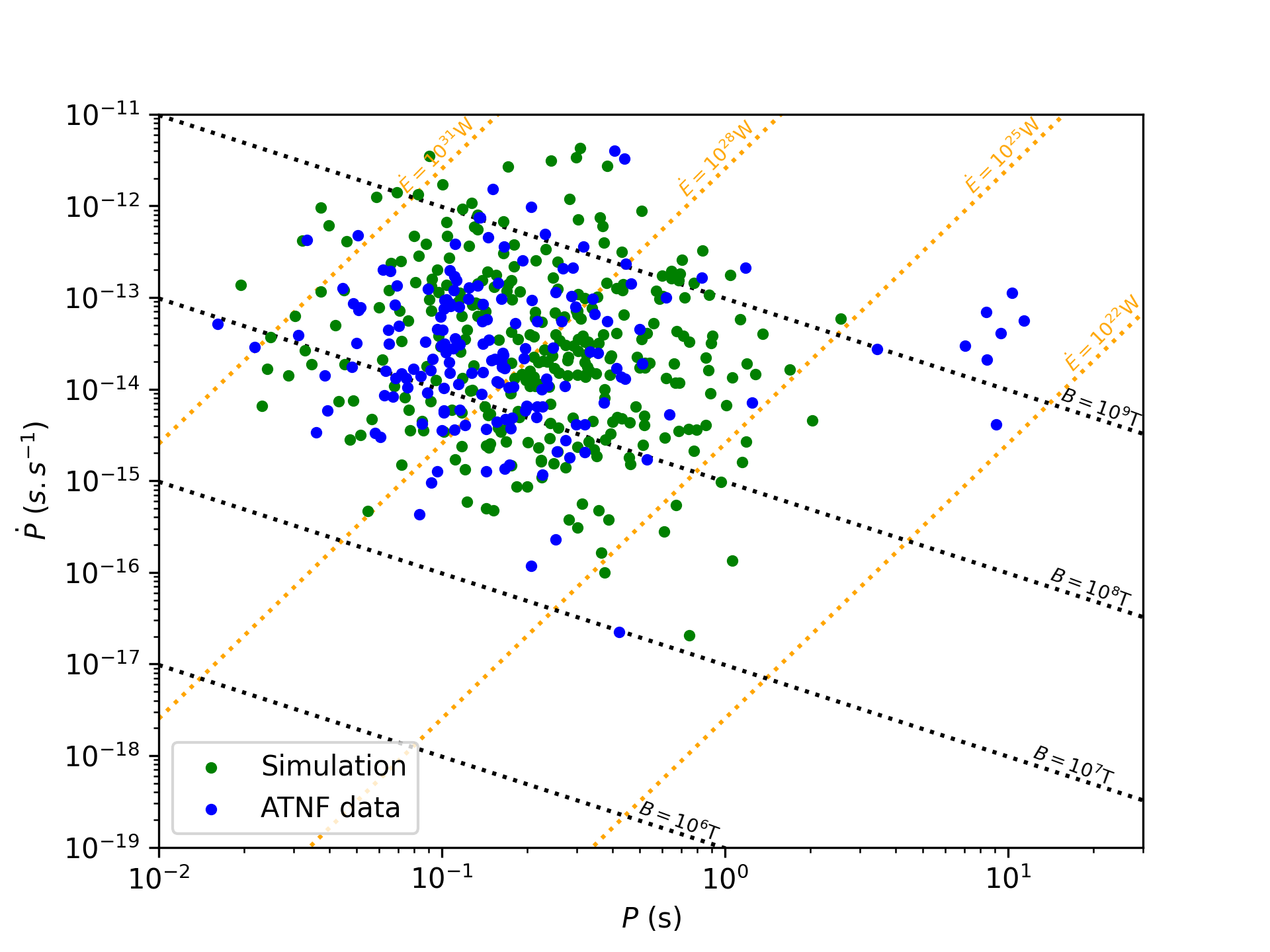}}
\caption{$P-\dot{P}$ diagram of the $\gamma$-only pulsars for both the simulations and the observations, with 1 million pulsars simulated.}
\label{PPdot_1M_gammaonly}
\end{figure}

Let us now focus on the $\gamma$-ray pulsar population. First, we note that the total number of pulsars detected is significantly larger in our sample (307 detections) compared to the observations (173 detections), when using the Fermi/LAT instrument sensitivity. However, the simulated $\gamma$-ray only and radio loud $\gamma$-ray pulsars are in the expected area of Fig.~\ref{PPdot_1M_gammaonly}, furthermore, the KS test gives a p-value of 0.52 and $\approx0$ for $\dot{P}$ and $P$ respectively. The null hypothesis can not be rejected for $\dot{P}$ but it is rejected for $P$. However, when the KS test is conducted on the $\gamma$-ray population only we get p-values of 0.03 and 0.14 for $\dot{P}$ and $P$ respectively, therefore here the null hypothesis is rejected for $\dot{P}$ but not for $P$. While for the radio loud $\gamma$-ray population p-values of 0.11 and $\approx0$ are obtained for $\dot{P}$ and $P$ respectively. Thus, we thought the low p-value obtained for $P$ for both population was probably because of the 7 high spin period radio loud $\gamma$-ray pulsars seen on Fig. \ref{PPdot_1M_gammaonly}, but even when conducting the KS test without these pulsars gave a p-value $\geq 0.05$ for $\dot{P}$ and well below $0.05$ for $P$.

\begin{figure}[h]
\resizebox{\hsize}{!}{\includegraphics{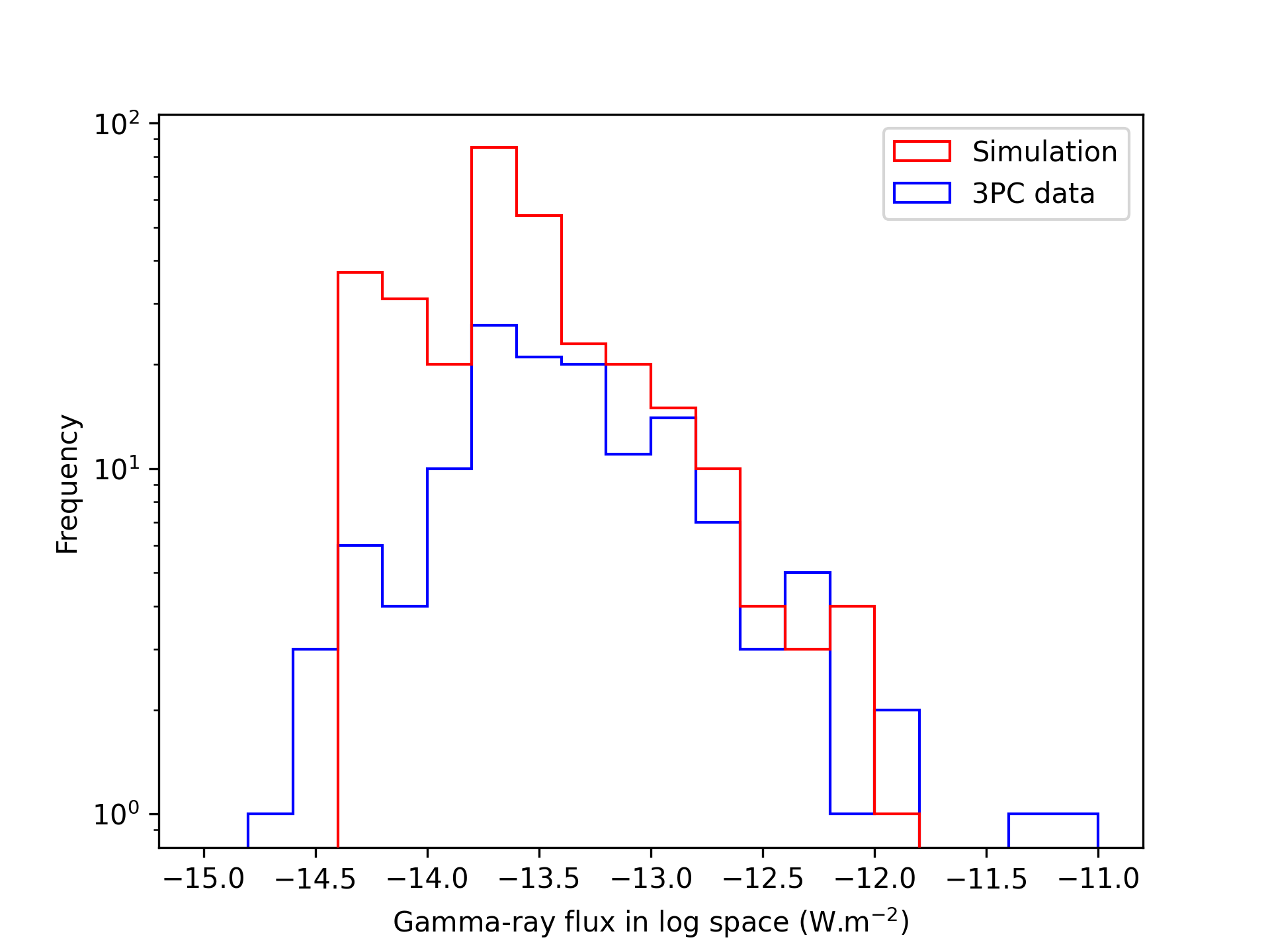}}
\caption{Distribution of the $\gamma$-ray flux of the simulated population along with the observations from the 3PC catalogue.}
\label{histo_fluxgamma}
\end{figure}

In addition, we also compare the $\gamma$-ray flux distribution of the simulated and observed populations. In Fig.\ref{histo_fluxgamma} we notice that both distributions peak at the same flux level around $10^{-13.7}$~W.m$^{-2}$. Globally, the number of pulsars in the simulation with a flux greater than $10^{-13.4}$~W.m$^{-2}$ is very close to the observations. However, between $10^{-14.4}$~W.m$^{-2}$ and $10^{-14}$~W.m$^{-2}$, there are much more pulsars in the simulation. This bi-modality in the simulated $\gamma$
-ray flux distribution is an artefact introduced by our abrupt condition on the threshold $F_{\rm min}$ depending on the radio detection of a pulsar and on a sufficiently low galactic latitude. However if we smoothly change $F_{\rm min}$ from $4\times10^{-15}$ to $16\times10^{-15}$ (the two $F_{\rm min}$ values used in this work), the bi-modality would almost disappear.

\begin{figure}[h]
\resizebox{\hsize}{!}{\includegraphics{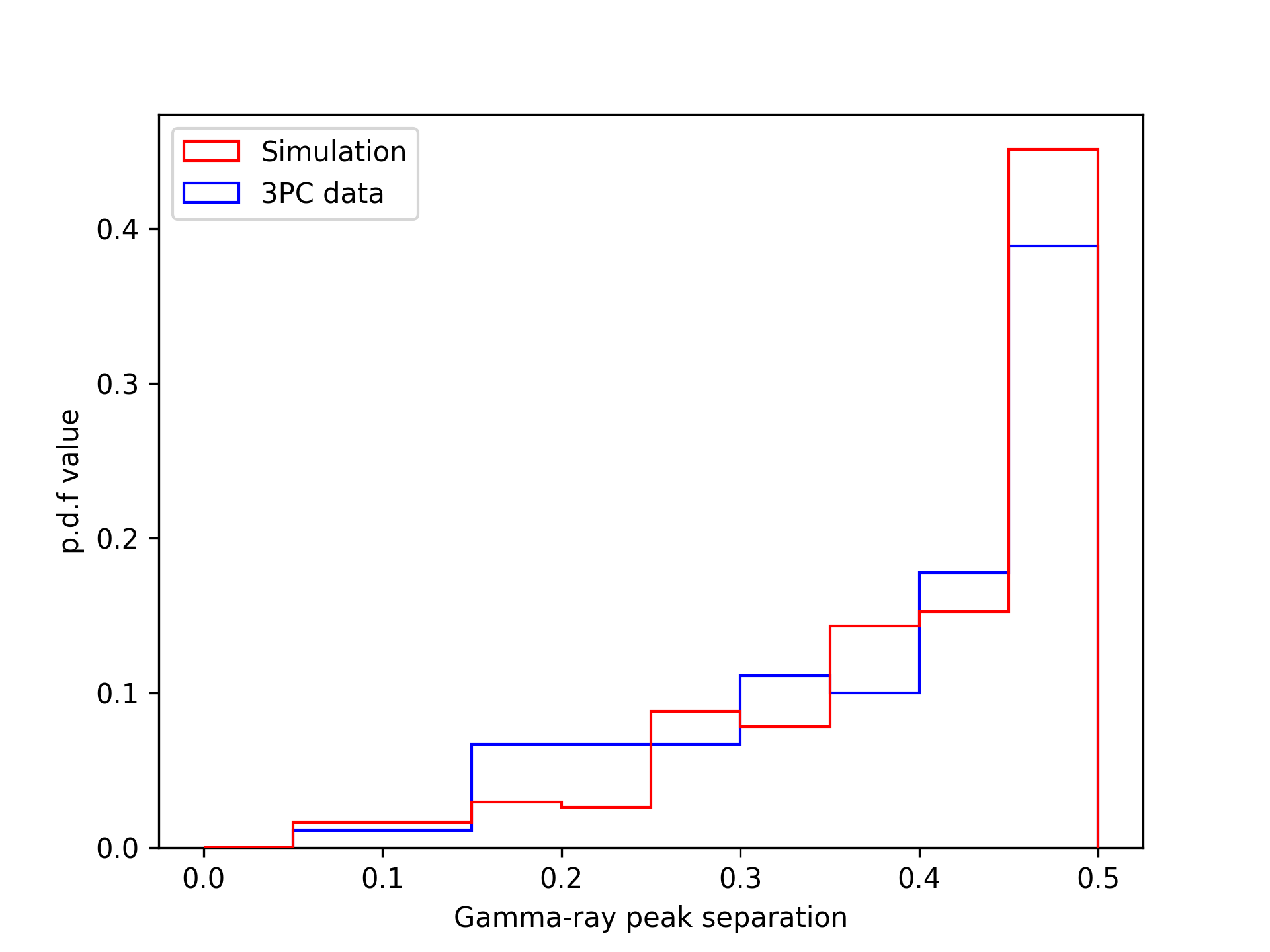}}
\caption{Distribution of the $\gamma$-ray peak separation of the simulated population along with the observations from the 3PC catalogue.}
\label{histo_peaksep}
\end{figure}

\modif{Finally, we compare the simulated and observed $\gamma$-ray light-curve peak separation $\Delta$ in Fig.~\ref{histo_peaksep}. The separation $\Delta$ is computed according to \citet{Petri2011} by
\begin{equation}
\cos\left(\pi \, \Delta\right) = |\cot\xi \cot\alpha | 
\end{equation}
where $\xi$ is the angle between the line of sight and the rotation axis and $\alpha$ the inclination angle. The distributions are normalized in order to obtain probability distribution functions (p.d.f) for both observations and simulations. The normalization is performed by dividing each distribution function by its total number of pulsars, observed or simulated. Moreover, $\Delta$ is constrained to the interval between 0 and 0.5, because we consider that peak separations~$\Delta$ larger than 0.5 can be scaled to $\Delta = 1 - \Delta$ by a phase shift, since the signal is periodic and represents the phase separation between two peaks and the definition of the first or main peak is arbitrary. Fig.~\ref{histo_peaksep} displays a remarkable similarity between the observed and simulated distributions of the $\gamma$-ray peak separation, meaning the striped wind model captures very well this geometric feature.}

The higher number of detected pulsars in the simulations could be related to the unidentified sources in the Fourth Fermi-LAT catalogue of $\gamma$-ray sources (4FGL). 
Since many unidentified Galactic sources are listed in this catalogue, many $\gamma$-ray pulsars might have been observed without being identified as such yet. Nonetheless, assuming that every unidentified $\gamma$-ray pulsar in this catalogue will be recognized as such by a future more sensitive instrument and that a significant difference between the number of simulated and observed pulsars subsists, this could hint to some shortcomings faced by the striped wind model. Let us mention some of these possible flaws of geometrical or physical nature. First the $\gamma$-ray beam opening angle could be smaller than the one used due to some plasma recollimation effect within the striped wind. This would reduce the number of detected pulsars. Second the luminosity function could also deviate from the prescription adopted in \eqref{eq:gamma_lum}. Another dependence on $B$ and $P$ or equivalently on $P$ and $\dot{P}$ could be introduced, maybe adding a third parameter like the cut-off energy to better describe the $\gamma$-ray luminosity. 
Furthermore, we compare our results with the PPS work of \citet{Pierbattistaetal}, where they simulated the young $\gamma$-ray pulsar population with 4 different models : Outer Gap (OG), Slot Gap (SG), Polar Cap (PC) and Outer Polar Cap (OPC) models. To summarize their findings, their best model is the OPC model, showing almost identical results as ours, $P-\dot{P}$ shape and number of detection, except that we detect $\gamma$-ray pulsars a little farther away from Earth. 


To conclude the part about $\gamma$-ray pulsars, we predict the number of pulsars that would be detected by a future instrument, ten times more sensitive than Fermi/LAT. We asked the question: what if we run another simulation with the same parameters as in Table~\ref{tabl_sim_para}, only changing the threshold of detection for $\gamma$-ray? This investigation could help in identifying the nature of the GeV excess in the Galactic centre.  
Therefore we decreased the instrument sensitivity~$F_{\rm min}$ by a factor~10: if the galactic latitudes of the pulsar is < 2°, then $F_{\rm min} = 0.4 \times 10^{-15}$ W.m$^{-2}$, and concerning blind searches we set $F_{\rm min} = 1.6 \times 10^{-15}$ W.m$^{-2}$.

\begin{figure}[h]
\resizebox{\hsize}{!}{\includegraphics{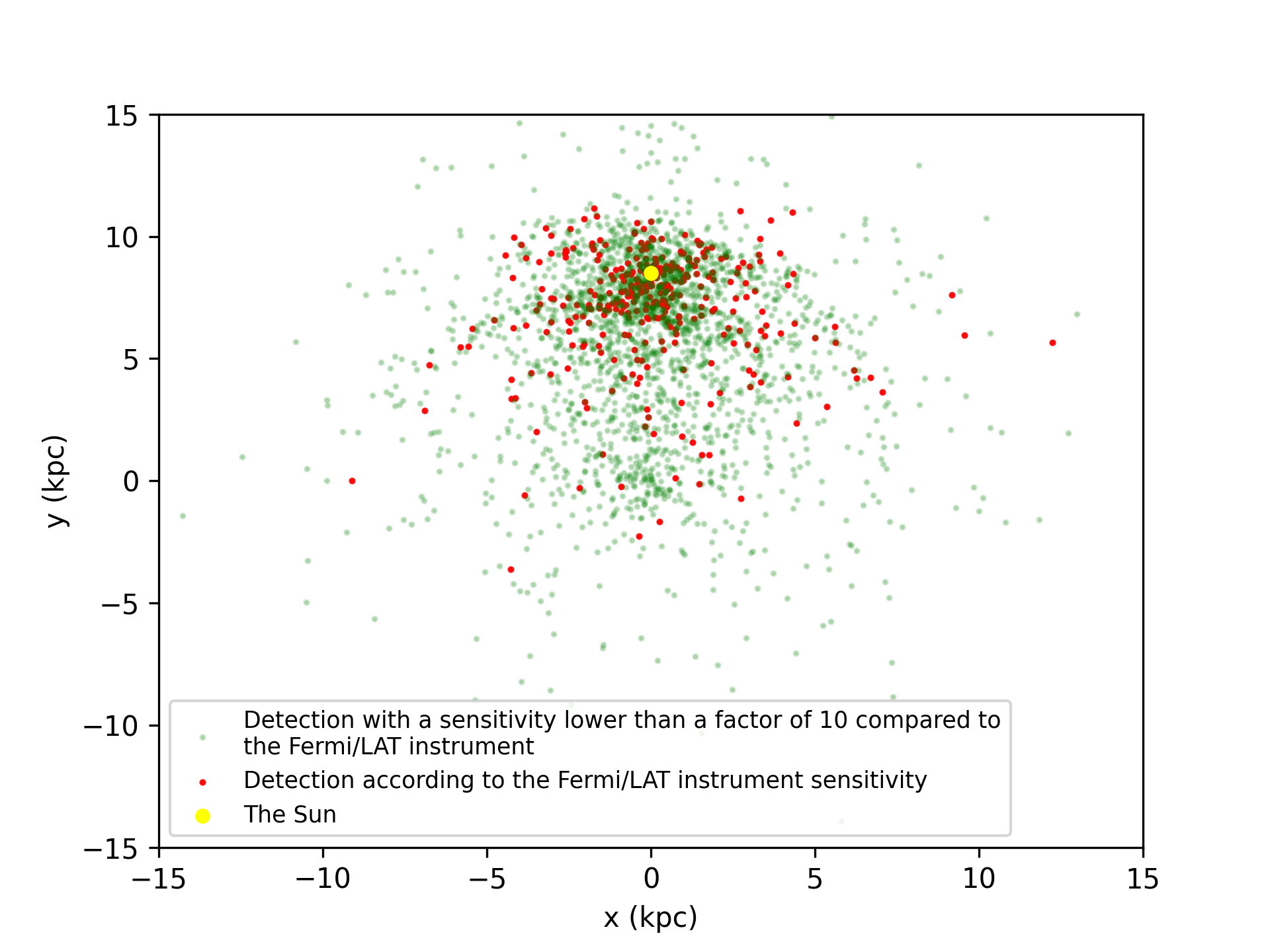}}
\caption{Spatial distribution of the simulated $\gamma$-ray detected pulsars, projected onto the Galactic plane for 1 million pulsars simulated, with a lower $F_{min}$.}
\label{XY_gamma_ab_LAT}
\end{figure}

The $\gamma$-ray detected pulsar positions in the X-Y plane of the Galaxy are shown in Fig.~\ref{XY_gamma_ab_LAT}. In this simulation, 2492~$\gamma$-ray pulsars are detected, in green, being $\gamma$-ray only or radio-loud $\gamma$-ray pulsars, among them only 350 (really close to the number of $\gamma$-ray and radio/$\gamma$-ray pulsars detected in Subsection \ref{sim_1M}) would have been detected if the same sensitivity as before was kept, in red in Fig.~\ref{XY_gamma_ab_LAT}. Therefore we detect 7~times more $\gamma$-ray pulsars compared to the number found with the previous sensitivity. Furthermore, the other 2142~pulsars, in green in Fig.~\ref{XY_gamma_ab_LAT}, detected with this increased sensitivity highlight several interesting features. We obviously detect more pulsars close to Earth, and also many more in the centre of the Galaxy. This result indicates that the GeV excess in the Galactic centre could be linked to pulsars not yet identified, while usually the origin of this GeV excess is attributed to self-annihilating dark matter particles \citep{Hooper22}. However, millisecond pulsars, not modeled in this work, could also contribute significantly to this GeV excess.

\section{Discussion} \label{S5}

For the initial spin period distribution \citet{Igo} suggested values for the mean period $P_{\text{mean}}$ in the range from approximately $57$~ms to $129$~ms and for its spread $\sigma_p$ the range from $0.45$ to $0.65$. The values $P_{\text{mean}}=129$~ms and $\sigma_p=0.45$ are best educated guesses. However further investigations on this work could benefit from an automated method to find the best set of values. Hence, we are currently working on an optimization algorithm to find the optimal initial parameters for the population synthesis, not only for the spin period, but for the whole parameter space presented in Table~\ref{tabl_sim_para}. 
Our optimization algorithm will help to constrain the properties of the Milky Way pulsar population by constraining parameters like the birth rate which is supposed to be between $1/33$~yr$^{-1}$ and $1/150$~yr$^{-1}$, see \citet{FaucherG}, \citet{Gullon14} or \citet{Johnston2017}.



In the main part of the paper, we abstain from showing outcomes stemming from simulations featuring a constant magnetic field. However we ran such simulations that clearly highlight the inferior performance of a simulation without magnetic field evolution as can be checked in appendix~\ref{AppA}. Moreover appendix~\ref{AppB} explores different parameter spaces for the PPS in order to meticulously scrutinizing whether the parameters showcased in the main part of the paper optimally encapsulate observational data. This work demonstrates that even if a substantial number of parameters are required for the PPS, they can be meaningfully constrained by such studies with reasonable accuracy. 

Furthermore, in Subsection \ref{gamma_results} we explored the results when the sensitivity in $\gamma$-ray is increased by a factor of 10. However we did not discuss improvements in the radio detection yet, while we run a simulation by increasing the sensitivity by a factor of 50 to match approximately the properties of SKA \citep{SKApaper}. The result of this simulation with an increased sensitivity led to 52276 pulsars detected (when simulating 1 million pulsars, with a death line), therefore approximately 31 times more than with the sensitivity of Parkes and Arecibo telescopes.

We emphasize that the death valley strongly depends on the chosen model. For instance the radio luminosity law plays a central role, we assumed that $F_r \propto \dot{E}^{\alpha}$ with $\alpha = 1/4 $ (see equation \eqref{eq:rad_lum}) according to \citet{Johnston2017} and \citet{Johnston20} where they found it to best fit to the data. Following their conclusion we decided not to choose $\alpha$ as another parameter. However it would impact the death valley, but this effect is not considered in our work. Furthermore, in other works like \citet{Graber} for instance, they do not explicitly use a death valley because they use a different radio luminosity law, meaning a different value for $\alpha$ and adjust the magnetic field decay time accordingly. We stress that this is just another method to deal with the bottom right corner of the $P-\dot{P}$ diagram, where pulsars are likely to stop emitting photons, since the death valley removes pulsars based on a combination of $P$ and $\dot{P}$. \citet{Graber} PPS is based on a phenomenological approach where the radio luminosity is tuned to conform to observations. In our approach, adding a death valley is based on a more physical ground related to the microphysics close to the neutron star polar cap.

\section{Summary} \label{S6}

Without the implementation of the Galactic potential, in the PPS of \citet{Dirson22}, the pulsars were moving in straight lines along one direction chosen randomly for each pulsar. This evolution of the proper motion was not realistic, even though the evolution model had already replicated a significant portion of the canonical pulsar population. However all the new improvements put into this PPS, meaning the Galactic potential, the effect of the ISM on the radio pulse profiles and the fact that the decay timescale of the magnetic field is taken randomly within a certain range, allow a more accurate reproduction of the population in the $P-\dot{P}$ diagram. 

The first results where we emulate 10~millions pulsars showed an excess of old pulsars detected in the simulation. In the $P-\dot{P}$ diagram we saw that this excess was associated to pulsars which lie below the death line, therefore, not creating pairs any more. With this finding we realized that it was imperative to implement the death line. The results obtained with the death line were then in better agreement with observations when plotting the most relevant distributions such as period and period derivative. 

Nonetheless, simulating 1~million pulsars allowed to show that with or without a death line the results were alike and therefore, in the case where the oldest simulated pulsar is $4.1\times10^7$~yr (its real age), a death line is not necessary to explain the observations. In this paper we showed two approaches to reproduce the observed canonical pulsar population, however when we consider the approach where the oldest pulsar simulated is only $4.1\times10^7$~yr we state that no older pulsars might \modif{be detected with our current telescopes} in the Galaxy, while the Milky Way is very much older than this \modif{and therefore has been forming pulsars since its beginning}. There is apparently no reason not to simulate pulsars that are \modif{older than $10^{8}$~yr}, and if there is one we do not know it, they become harder to detect because of their wide pulse profile with their age, but some \modif{could} still \modif{be} detectable. If this situation is the reality then it means that the death line is not necessary in pulsar population synthesis to explain observations. When the simulation is ran with 10 millions pulsars, the death line becomes a necessity. However this simulation gives results less similar to observations.

It was also shown that old pulsars with age larger than 10~Myr have a tendency to lose their alignment between rotation axis and velocity, a direct effect of the Galactic potential as shown by the observations of \citet{Noutsosetal}, but it is the first time it is done in a simulation.

The striped wind model seems to reproduce the $\gamma$-ray population of canonical pulsars well according to the p-value of the KS test greater than 0.05 for $\dot{P}$. However the KS test gives a p-value lower than 0.05 for $P$, but there are several observed radio-loud $\gamma$-ray pulsars that the simulation struggles to reproduce (not only the ones with high $P$) and might explain this low p-value. However in the simulation a higher number of these pulsars are detected compared to observations. It probably indicates that a part of the unidentified sources from the 4FGL catalogue are pulsars. Testing an instruments with improved sensitivity for the detection of $\gamma$-ray pulsars lead us to the conclusion that part of the GeV excess in the Galactic centre could be young $\gamma$-ray pulsars. Moreover the assumption that millisecond pulsars could also contribute to this excess will be checked once a PPS model for millisecond pulsars will be available. The most common picture suppose that only millisecond pulsars are responsible for this excess, while we demonstrated that young $\gamma$-ray pulsars could also play a significant role.

In summary, the best parameters found for this pulsar population synthesis are for $P_{\text{mean}}=129$~ms, $\sigma_p=0.45$ for the log-normal spin period distribution at birth, $B_{\text{mean}}=2.75 \times 10^8$~T, $\sigma_b=0.5$ for the magnetic field distribution at birth and for a birth rate of 1/41~yr$^{-1}$. Ultimately, this work provides a simulated population of pulsars within the Milky Way more similar to the observations compared to \citet{Johnston20} and \citet{Dirson22}, the two most recent population synthesis considering both radio and $\gamma$-ray emission.  
Moreover, such investigations are useful to predict the detection rate of future radio surveys such as SKA or NenuFAR.

The PPS method developed in this paper can be extended to other pulsar populations in the Milky Way such as millisecond pulsars and magnetars. The magnetars, for instance, would need another distribution at birth for their magnetic field, taking into account the emission in X-rays and having a faster decay for the magnetic field. The magnetars are located just above the canonical pulsars in the $P-\dot{P}$ diagram, at the top right of the diagram. A more complex population is the recycled or millisecond pulsars, that were "brought back to life" by accreting matter from a companion star. This accretion mechanism spins up the neutron stars moving them from the graveyard of pulsars to the bottom left of the $P-\dot{P}$ diagram, crossing again the death line in the opposite sense and allowing them to revive by creating electron-positron pair cascades again. However, they are difficult to model because their evolution scenario must take into account their accretion phase and spin-up. We currently work on both these populations to reproduce them in a PPS based on this work.

\begin{acknowledgements}
This work has been supported by the grant ANR-20-CE31-0010. We acknowledge the High Performance Computing Centre of the University of Strasbourg for supporting this work by providing scientific support and access to computing resources. D.M. acknowledges the support of the Department of Atomic Energy, Government of India, under project No. 12-R\&DTFR-5.02-0700.
\end{acknowledgements}
\bibpunct{(}{)}{;}{a}{}{,}
\bibliographystyle{aa}

\begin{appendix}

\section{The integration scheme used} \label{AppC}

\subsection{The scheme}
One of the new aspect in this work is to solve the equation of motion~\eqref{move_equation} in order to follow more realistically the motion of our sample of pulsars. To achieve this objective we implemented a fourth order integration scheme called Position Extended Forest Ruth-Like (PEFRL). This algorithm was chosen because of its high order allowing a high precision on the trajectory of the pulsar. In addition, it is a symplectic algorithm, meaning that it conserves the total energy of the pulsar (kinetic plus potential), minimizing the error even if the integration is done for a long time with any orbits. It works in the same way as a Verlet scheme of second order, the velocities and coordinates are shifted separately but in the end they get synchronized. The different steps of the algorithm are as follows
\begin{subequations}
\begin{align}
\vec{r}^{n+\frac{1}{5}} & = \vec{r}^n + \xi h \vec{v}^n \\
\vec{v}^{n+\frac{1}{4}} & = \vec{v}^n - (1-2\lambda) \frac{h}{2} \vec{\nabla} \Phi(\vec{r}^{n+\frac{1}{5}}) \\
\vec{r}^{n+\frac{2}{5}} & = \vec{r}^{n+\frac{1}{5}} + \chi h \vec{v}^{n+\frac{1}{4}} \\
\vec{v}^{n+\frac{2}{4}} & = \vec{v}^{n+\frac{1}{4}} - \lambda h \vec{\nabla} \Phi(\vec{r}^{n+\frac{2}{5}}) \\
\vec{r}^{n+\frac{3}{5}} & = \vec{r}^{n+\frac{2}{5}} +(1-2(\chi+\xi)) h \vec{v}^{n+\frac{2}{4}} \\
\vec{v}^{n+\frac{3}{4}} & = \vec{v}^{n+\frac{2}{4}} - \lambda h \vec{\nabla} \Phi(\vec{r}^{n+\frac{3}{5}}) \\
\vec{r}^{n+\frac{4}{5}} & = \vec{r}^{n+\frac{3}{5}} + \chi h \vec{v}^{n+\frac{3}{4}} \\
\vec{v}^{n+1} & = \vec{v}^{n+\frac{3}{4}} - (1-2\lambda) \frac{h}{2} \vec{\nabla} \Phi(\vec{r}^{n+\frac{4}{5}}) \\
\vec{r}^{n+1} & = \vec{r}^{n+\frac{4}{5}} + \xi h \vec{v}^{n+1} 
\end{align}    
\end{subequations} 
here $\vec{r}$ is the position vector, $\vec{v}$ is the velocity vector, $h$ is the time step, $\vec{\nabla} \Phi$ is the gradient of the potential and $n$ is the time at which we compute the position or the velocity. $\lambda$, $\xi$ and $\chi$ are constants that are tabulated in Table~\ref{tabl_const_PEFRL}, see for instance \citet{PEFRL}. 

\begin{table}[h]
\caption{Value of the constants used for the PEFRL scheme.}
\label{tabl_const_PEFRL} 
\centering 
\begin{tabular}{c c} 
\hline\hline 
Parameters & Values \\
\hline 
$\lambda$ & -0.212341831062605 \\ 
$\xi$ & 0.178617895844809 \\
$\chi$ & -0.06626458266981849  \\
\hline 
\end{tabular}
\end{table}

\subsection{Precision of the PEFRL algorithm}

\begin{figure}[h]
\resizebox{\hsize}{!}{\includegraphics{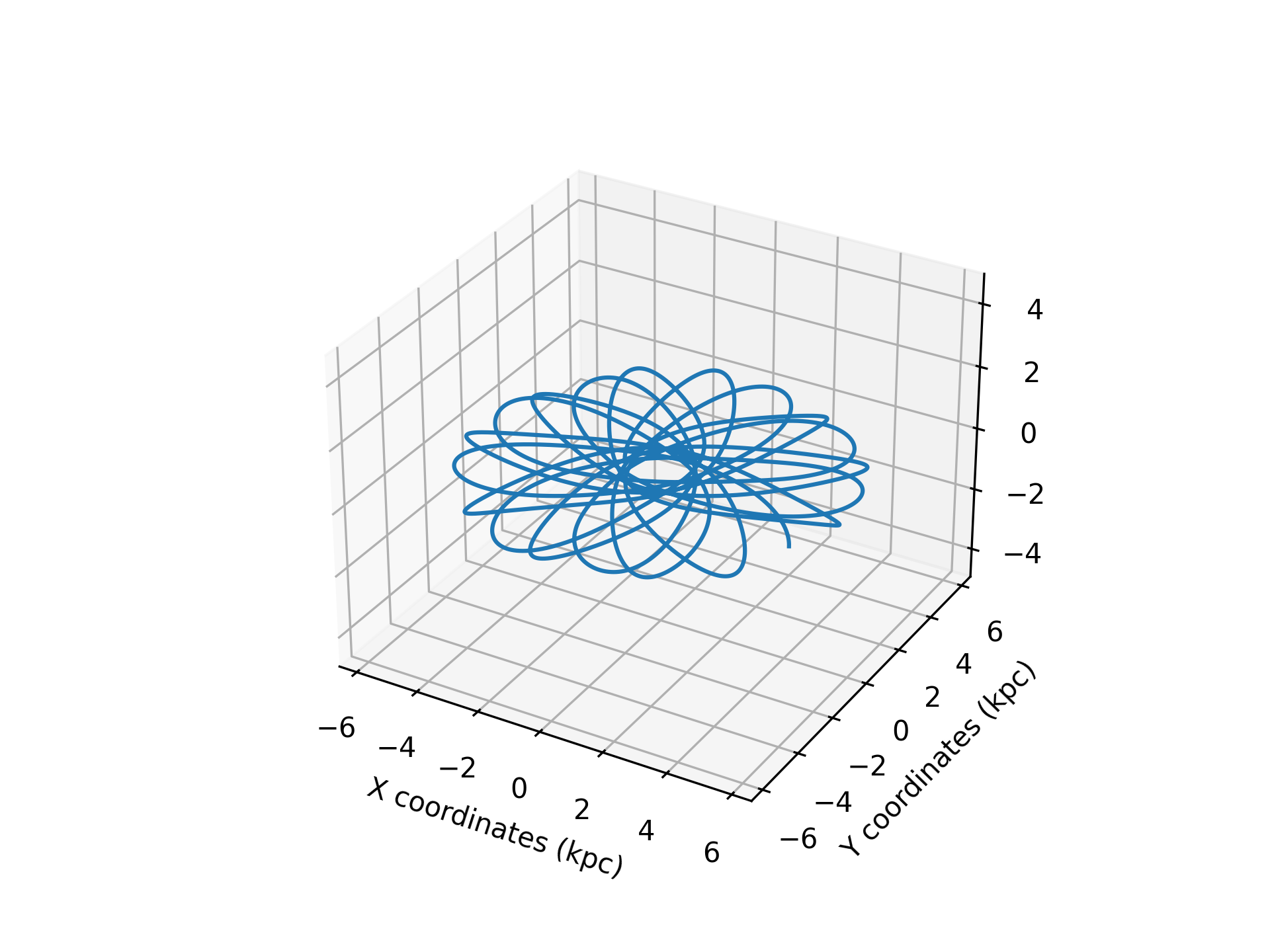}}
\caption{Trajectory of a single pulsar in the Galactocentric frame.}
\label{Pulsar_traj}
\end{figure}

Before implementing the gravitational potential and the integration scheme in the whole synthesis, the precision of the trajectory of a single pulsar in the Milky Way was checked. The initial conditions for the pulsar are the following : $x= 5.1$~kpc, $y = -1.9$~kpc, $z = 0.05$~kpc, $v_x = -47.8$~km/s $v_y = 97.0$~km/s, and $v_z$ = 22.6~km/s. Those initial conditions were chosen randomly, the only constrain being to have the pulsar in the Galaxy with realistic coordinates on the one hand and with realistic speeds for a pulsar on the other hand. Furthermore, the total integration time is 1.25~Gyr and the time step used is $10^5$~yr. In order to check the precision of the scheme, the total energy of the pulsar, sum of the kinetic plus potential energy, that should be conserved is computed as
\begin{equation} \label{tot_energy}
E_{tot} = \frac{v_x^2}{2} + \frac{v_y^2}{2} + \frac{v_z^2}{2} + \Phi_{tot} .
\end{equation}
The trajectory obtained after the integration with the PEFRL algorithm is shown in Fig.~\ref{Pulsar_traj}. This trajectory shows that the pulsar, with these initial conditions, remains bound to the Galaxy. The pulsar here takes approximately 201.3 Myr to complete an orbit in the Galaxy. The time step chosen is significantly smaller than the orbital period ($h/P_{\rm orb} = 0.0005$), and we obtain approximately 6 orbits by integrating the trajectory during 1.25 Gyr. \cite{BB21} showed several trajectories in their paper, the one obtained here seems coherent compared to the different trajectories that they got.

\begin{figure}[h]
\resizebox{\hsize}{!}{\includegraphics{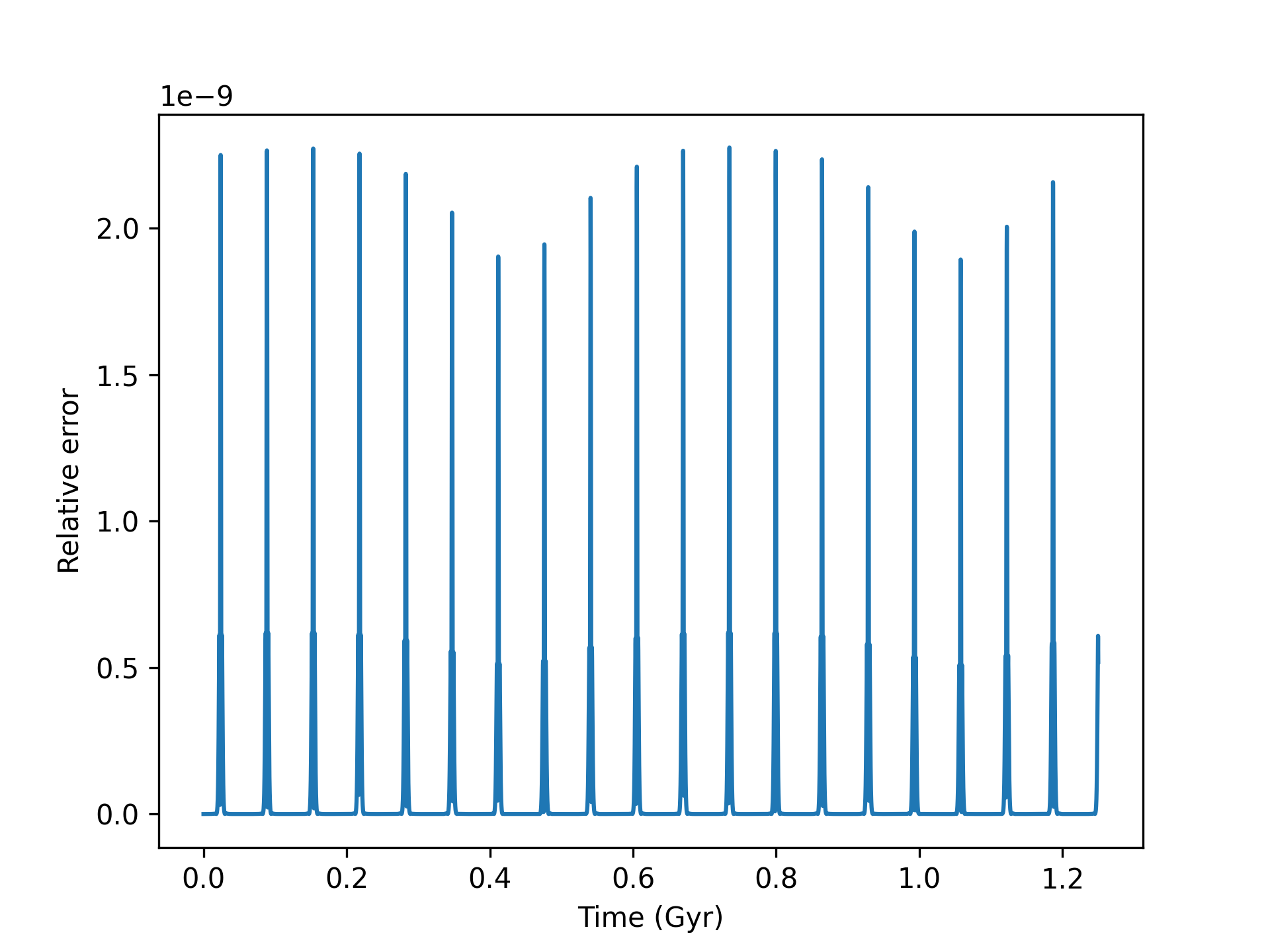}}
\caption{Relative error on the energy of a single pulsar in the Galactocentric frame.}
\label{Pulsar_error}
\end{figure}

Fig.~\ref{Pulsar_error} shows the relative error on the total energy. It is indeed almost constant along the whole integration of the trajectory. The maximum relative error on the energy is of the order of $10^{-9}$, a very good accuracy for this trajectory. Globally, when a single pulsar has its trajectory integrated in the PPS, the relative error is at maximum 0.001 and the minimum seen was $10^{-14}$. 

\section{Results for a constant magnetic field} \label{AppA}

Depending on the PPS, the magnetic field will be considered constant \citep{Johnston2017,FaucherG} or decaying \citep{Gullon14,Dirson22,Jawor22}. The goal is here is to show the results with a constant magnetic field in order to see which model is better between the decaying magnetic field and the constant magnetic field. 

\begin{figure}[h]
\resizebox{\hsize}{!}{\includegraphics{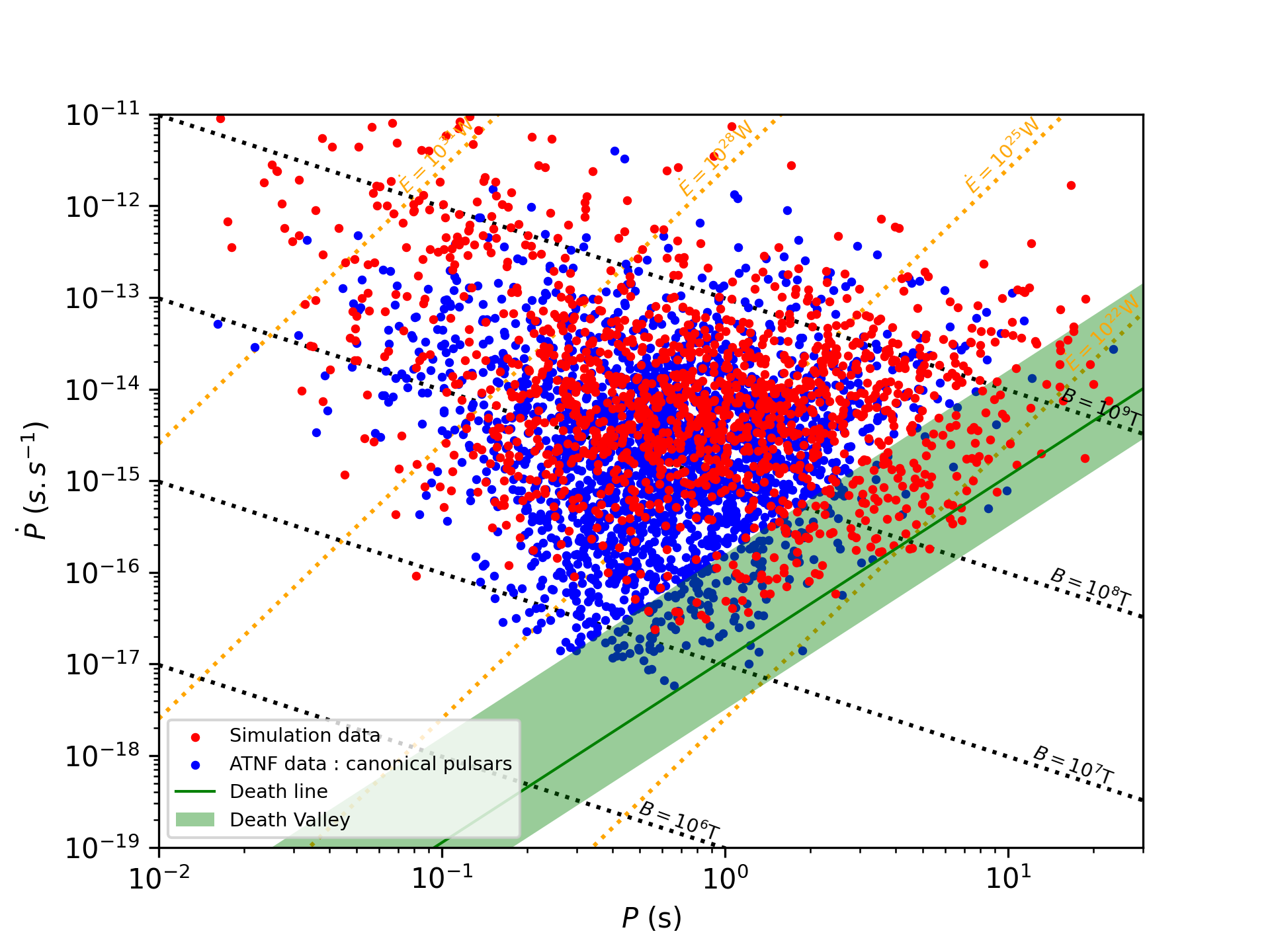}}
\caption{$P-\dot{P}$ diagram of the simulated population, along with the observations, with the death line implementation and a constant magnetic field for each pulsar.}
\label{P_Pdot_cst_B}
\end{figure}

Hence, this simulation was run with \modif{30}~million pulsar simulated \modif{(meaning the oldest pulsar simulated is $1.23\times10^{9}$~yr)}, \modif{with} a death line and without a decaying magnetic field. The KS test gave p-values of \modif{$1.03\times10^{-22}$ and $3.38\times10^{-51}$ for $P$ and $\dot{P}$ respectively}, therefore the null hypothesis is rejected for both $P$ and $\dot{P}$. \modif{Here we simulate more pulsars than in the main part of the paper, since if we simulate with the same number of pulsars, it seems the whole $P-\dot{P}$ diagram is not covered. Therefore if a constant magnetic field was the best way to reproduce the canonical population, it means we would detect much older pulsars than with the magnetic field decay prescription.} The parameters used are in Table \ref{tabl_sim_para}. It gives results less close to the observations than with a decaying magnetic field. As can be seen in Fig. \ref{P_Pdot_cst_B}, too many pulsars from the simulation appear on the top right of the diagram and almost none at the bottom. In Table \ref{tabl_nbpulsar_cst_b}, we note a total of pulsars which is smaller from what we found for all of our previous results in Sect. \ref{S4}. We could expect to reproduce equally well the population of observed pulsars with a constant magnetic field, however compared to the study of \citet{Johnston2017} where they used this assumption for the magnetic field, we do not evolve the braking index. In our simulations, we have $\dot{\alpha} < 0$, $\dot{B} < 0$ and $\dot{n} = 0$ whereas in their simulation $\dot{\alpha} < 0$, $\dot{B} = 0$ and $\dot{n} \neq 0$. Having either the braking index or the magnetic field evolve, allows to reproduce well the population of pulsars.

\begin{table}[h]
\caption{Number of pulsars detected for the simulation with constant magnetic field for each pulsar and the death line implementation.}
\label{tabl_nbpulsar_cst_b} 
\centering 
\begin{tabular}{c c c c c} 
\hline\hline 
log($\dot{E}$) (W) & $N_{tot}$ & $N_r$ & $N_g$ & $N_{rg}$\\
\hline 
>31 & 42 & 0 & 29 & 13\\ 
>28 & 213 & 10 & 127 & 76\\
Total & 1473 & 1075 & 215 & 183\\
\hline 
\end{tabular}
\end{table}

\section{Influence of the different parameters} \label{AppB}

This appendix shows that the parameters chosen in the main part of this paper are meticulously picked and that the uncertainties brought by the large number of parameters is manageable. Multiple simulations were conducted by varying parameter values for the initial distributions, demonstrating that deviations from the optimal values lead to poor results. The simulations were run with in the same conditions as subsection~\ref{sim_1M}, with only one parameter changing between the different sets of simulations and without implementing the ISM, because it would be too time consuming to perform all these simulations. As a consequence in the tables from \ref{tabl_dval_pmean} to \ref{tabl_dval_BR}), the p-values are low, while when including the ISM effect they become larger. Therefore, the results and conclusions presented in this section are based on this simplification, which was primarily used to guide the choice of parameters. While these assumptions have helped to identify efficient parameters, it is important to stress that the proposed values are not necessarily optimal. They serve primarily as a foundation for refining the parameters under more realistic conditions. That is why before validating a parameter we checked that the p-values were satisfactory when including the ISM. We found each time a higher p-values.

In order to get an idea on a good range of values for the parameters, the d-values and p-values obtained thanks to the KS test are shown in Table \ref{tabl_dval_pmean}, \ref{tabl_dval_Bmean}, \ref{tabl_dval_sigb}, \ref{tabl_dval_sigp} and \ref{tabl_dval_BR}. For each parameter, 10 simulations were performed in order to get a mean and a standard deviation for the d-values and p-values associated. These simulations helped finding the best values for the parameters by considering the parameter is well constrained when its d-values for both distribution of $P$ and $\dot{P}$ are below or close to 0.05 (meaning 5\% similarity between the observed and simulated distributions), especially if the previous and next parameter values have higher d-values and lower p-values. As usual p-values above 0.05 means that the null hypothesis can not be rejected, however as explained above, without the ISM effect taken into account the p-values for the $\dot{P}$ distribution do not get too high here. Even though it is not an optimisation (and therefore the results can be improved on that point), it allows us to conclude that $P_{\rm mean}$ must be between 120~ms and 135~ms, with 129~ms as best value. $B_{\rm mean}$ must be between $2.25\times10^8$ T and $2.8\times10^8$ T, with $2.75\times10^8$ T as best value. $\sigma_b$ is well constrained between 0.49 and 0.51, with 0.5 as best value. While $\sigma_p$ must be between 0.42 and 0.47, with 0.45 as best value. Finally, the birth rate must be between 1/37~yr$^{-1}$ and 1/45~yr$^{-1}$, with 1/41~yr$^{-1}$ as best value.

\begin{table*}[h]
\caption{D-value for different $P_{\rm mean}$.}
\label{tabl_dval_pmean} 
\centering 
\begin{tabular}{c c c c c} 
\hline\hline 
$P_{\rm mean}$ (in ms) & d-value ($\dot{P}$) & d-value ($P$) & p-value ($\dot{P}$) & p-value ($P$)\\
\hline 
90 & $0.072 \pm 0.0073$ & $0.058 \pm 0.0080$ & $0.0091 \pm 0.021$ & $0.038 \pm 0.049$\\ 
100 & $0.065 \pm 0.0069$ & $0.050 \pm 0.0064$ & $0.012 \pm 0.0089$ & $0.22 \pm 0.0086$  \\
110 & $0.065 \pm 0.0069$ & $0.038 \pm 0.0074$ & $0.0033\pm 0.0022$ & $0.32 \pm 0.0076$\\
120 & $0.055 \pm 0.0071$ & $0.030 \pm 0.0047$ & $0.0049 \pm 0.0060$ & $0.47 \pm 0.25$\\
129 & $0.056 \pm 0.0074$ & $0.028 \pm 0.0058$ & $0.010 \pm 0.014$ & $0.22 \pm 0.19$\\
135 & $0.054 \pm 0.0077$ & $0.033 \pm 0.0077$ & $0.012 \pm 0.015$ & $0.091 \pm  0.0095$\\
150 & $0.057 \pm 0.0078$ & $0.039 \pm 0.0093$ & $0.011 \pm 0.024$ & $0.013 \pm 0.0074$\\
\hline 
\end{tabular}
\end{table*}

\begin{table*}[h]
\caption{D-value for different $B_{\rm mean}$.}
\label{tabl_dval_Bmean} 
\centering 
\begin{tabular}{c c c c c} 
\hline\hline 
$B_{\rm mean}$ (in T) & d-value ($\dot{P}$) & d-value ($P$) & p-value ($\dot{P}$) & p-value ($P$)\\
\hline \ 
$1\times10^8$ & $0.11 \pm 0.0082$ & $0.20 \pm 0.0067$ & $4.4\times10^{-16} \pm 1.2\times10^{-15}$ &$8.9\times10^{-40}\pm2.4\times10^{-39}$\\
$2\times10^8$ & $0.042 \pm 0.0046$ & $0.066 \pm 0.0070$ & $0.033\pm 0.043$ & $ 0.0058\pm 0.0072$\\
$2.25\times10^8$ & $0.046 \pm 0.0045$ & $0.045 \pm 0.0061$ & $0.011 \pm 0.019$& $0.12 \pm 0.076$\\
$2.75\times10^8$ & $0.056 \pm 0.0074$ & $0.028 \pm 0.0058$ & $0.010 \pm 0.014$ & $0.22 \pm 0.19$\\
$2.8\times10^8$ & $0.058 \pm 0.0091$ & $0.030 \pm 0.0073$ & $0.0074\pm 0.0055$ & $0.20 \pm 0.16$\\
$3\times10^8$ & $0.065 \pm 0.011$ & $0.043 \pm 0.0077$ & $8.7\times10^{-4}\pm 9.0\times10^{-4}$ & $0.030 \pm 0.050$\\
$3.25\times10^8$ & $0.076 \pm 0.0093$ & $0.048 \pm 0.0010$ & $0.0040 \pm 0.0074$ & $0.0019 \pm 0.0034$\\
\hline 
\end{tabular}
\end{table*}

\begin{table*}[h]
\caption{D-value for different $\sigma_b$.}
\label{tabl_dval_sigb} 
\centering 
\begin{tabular}{c c c c c} 
\hline\hline 
$\sigma_b$ & d-value ($\dot{P}$) & d-value ($P$) & p-value ($\dot{P}$) & p-value ($P$)\\
\hline  
0.45 & $0.077 \pm 0.0067$ & $0.051 \pm 0.013$ & $7.2\times10^{-4}\pm 0.0011$& $0.0042\pm 0.0078$\\
0.48 & $0.062 \pm 0.0088$ & $0.038 \pm 0.0123$ & $0.0063\pm 0.0071$ & $0.0063\pm 0.0071$\\
0.49 & $0.065 \pm 0.0097$ & $0.030 \pm 0.0050$ & $0.059 \pm 0.057 $& $0.16 \pm 0.02437 $\\
0.5 & $0.056 \pm 0.0074$ & $0.028 \pm 0.0058$ & $0.010 \pm 0.014$ & $0.22 \pm 0.19$\\
0.51 & $0.056 \pm 0.0073$ & $ 0.031 \pm 0.0070$ & $0.0076\pm 0.0067$ & $0.30 \pm 0.18$\\
\hline 
\end{tabular}
\end{table*}

\begin{table*}[h]
\caption{D-value for different $\sigma_p$.}
\label{tabl_dval_sigp} 
\centering 
\begin{tabular}{c c c c c} 
\hline\hline 
$\sigma_p$ & d-value ($\dot{P}$) & d-value ($P$) & p-value ($\dot{P}$) & p-value ($P$)\\
\hline  
0.4 & $0.055 \pm 0.0024 $& $0.024 \pm 0.0056$ & $0.0036 \pm 0.0050 $ & $0.57 \pm 0.24$\\
0.42 & $0.054 \pm 0.0055$ & $ 0.028 \pm 0.0051 $ & $0.0050 \pm 0.0048 $ & $0.38 \pm 0.23 $\\
0.45 & $0.056 \pm 0.0074$ & $0.028 \pm 0.0058$ & $0.010 \pm 0.014$ & $0.22 \pm 0.19$\\
0.47 & $0.053 \pm 0.0049$ & $0.034 \pm 0.0092$ & $0.0096 \pm 0.015$ & $0.26 \pm 0.21$\\
0.48 & $ 0.053 \pm 0.0045 $ & $0.037 \pm 0.0084 $ & $0.0060 \pm 0.0050 $ & $0.17 \pm 0.18 $\\
0.52 & $0.053 \pm 0.0049$ & $0.043 \pm 0.011$ & $0.0082 \pm 0.012$& $0.11 \pm 0.14$\\
\hline 
\end{tabular}
\end{table*}

\begin{table*}[h]
\caption{D-value for different birth rate (BR).}
\label{tabl_dval_BR} 
\centering 
\begin{tabular}{c c c c c} 
\hline\hline 
BR (in yr~$^{-1}$) & d-value ($\dot{P}$) & d-value ($P$) & p-value ($\dot{P}$) & p-value ($P$)\\
\hline 
1/33 & $0.061 \pm 0.010$ & $0.031 \pm 0.0064$ & $ 0.013\pm 0.014$ & $0.49 \pm 0.17$\\ 
1/37 & $0.056 \pm 0.0085$ & $0.029 \pm 0.0064$ & $0.0086 \pm 0.0095$ & $0.48 \pm 0.21$\\
1/41 & $0.056 \pm 0.0074$ & $0.028 \pm 0.0058$ & $0.010 \pm 0.014$ & $0.22 \pm 0.19$\\
1/45 & $0.058 \pm 0.0051$ & $0.030 \pm 0.0056$ & $0.0055 \pm 0.0042$ & $0.18 \pm 0.14 $\\
1/70 & $0.069 \pm 0.0083$ & $0.034 \pm 0.0116$ & $4.3\times10^{-4} \pm 0.0012$& $0.048 \pm 0.074$\\
1/150 & $0.12 \pm 0.016$ & $0.041 \pm 0.0067$ & $5.9\times10^{-12} \pm 1.2\times10^{-11}$ & $0.035 \pm 0.091 $\\
\hline 
\end{tabular}
\end{table*}

\end{appendix}
\end{document}